\documentclass[letter]{aa} % for the letters 
\usepackage{txfonts}
\usepackage{graphicx}% Include figure files
\usepackage[caption=false]{subfig}
\usepackage{dcolumn}% Align table columns on decimal point
\usepackage{bm}% bold math
\usepackage{tikz}
\usetikzlibrary{chains,shapes.multipart}
\usetikzlibrary{shapes,calc}
\usetikzlibrary{arrows,snakes,automata,backgrounds,petri,positioning}

\tikzstyle{targetsearchstep} = [rectangle, minimum size=1.25cm,
align=center, draw=black, fill=white]
\usepackage{mathtools}

\DeclareMathAlphabet{\mathpzc}{OT1}{pzc}{m}{it}

\DeclarePairedDelimiterX{\infdivx}[2]{(}{)}{#1\;\delimsize\|\;#2}

\usepackage{resizegather}
\usepackage{textcomp}
\usepackage{verbatim}
\usepackage{xspace}

\newcommand\numberthis{\addtocounter{equation}{1}\tag{\theequation}}
\newcommand{\borg}{{\textsc{borg}}\xspace}

\newcommand{\Mpch}{{$h^{-1}$~Mpc~}}

\newcommand{\mvec}[1]{\bm{#1}}

\usepackage{xcolor}
\definecolor{blue}{HTML}{1F77B4}
 \definecolor{green}{HTML}{2CA02C}
\usepackage[%
  bookmarks=true,
  colorlinks,
  linkcolor=blue,
  urlcolor=blue,
  citecolor=blue,
  plainpages=false,
  pdfpagelabels,
  final,
  breaklinks=true
]{hyperref}

\usepackage[switch]{lineno}
\begin{document}

   \title{Optimal machine-driven acquisition of future cosmological data}% Force line breaks with \\

    \author{Andrija Kostić
        \inst{1,2}\thanks{akostic@mpa-garching.mpg.de}
        \and
        Jens Jasche \inst{3}\thanks{jens.jasche@fysik.su.se}
        \and
        Doogesh Kodi Ramanah \inst{4}\thanks{ramanah@nbi.ku.dk}
        \and 
        Guilhem Lavaux \inst{5}\thanks{guilhem.lavaux@iap.fr}
    }
   \institute
   {
       Max-Planck-Institut für Astrophysik, Karl-Schwarzschild-Str. 1, 85748 Garching, Germany
    \and
        Ludwig Maximilians University, Geschwister-Scholl-Platz 1, 80539 München, Germany
    \and  
        The Oskar Klein Centre for Cosmoparticle Physics, Department of Physics, Stockholm University, AlbaNova, Stockholm, SE-106 91, Sweden
    \and
        DARK, Niels Bohr Institute, University of Copenhagen, Jagtvej 128, 2200 Copenhagen, Denmark
    \and
        Sorbonne Universit\'e, CNRS, UMR 7095, Institut d'Astrophysique de Paris, 98 bis bd Arago, 75014 Paris, France
    }

  \abstract{
  We present maps classifying regions of the sky according to their information gain potential as quantified by the Fisher information. These maps can guide the optimal retrieval of relevant physical information with targeted cosmological searches. Specifically, we calculate the response of observed cosmic structures to perturbative changes in the cosmological model and chart their respective contributions to the Fisher information. Our physical forward modeling machinery transcends the limitations of contemporary analyses based on statistical summaries to yield detailed characterizations of individual 3D structures. We demonstrate this using galaxy counts data and showcase the potential of our approach by studying the information gain of the Coma cluster. We find that regions in the vicinity of the filaments and cluster core, where mass accretion ensues from gravitational infall, are the most informative about our physical model of structure formation in the Universe. Hence, collecting data in those regions would be most optimal for testing our model predictions. The results presented in this work are the first of their kind and elucidate the inhomogeneous distribution of cosmological information in the Universe. This study paves a new way forward to perform efficient targeted searches for the fundamental physics of the Universe, where search strategies are progressively refined with new cosmological data sets within an active learning framework.}

   \maketitle
%
%-------------------------------------------------------------------
\section{Introduction}

``\textit{Where in the Universe should we look to learn something new?}'' This is one of the most outstanding questions asked by researchers of all times. It reflects the wish to confirm our current physical understanding of the world and find new evidence to shift prevailing scientific paradigms. The latest advances in cosmological observations and data analysis now unlock the opportunity to perform machine-aided targeted searches for cosmological physics across the Universe. This has become feasible due to the availability of large-scale inferences that are informed by physics and causality of the cosmic large-scale structures and their evolution with time.   

State-of-the-art cosmological surveys gather scientific information from tracers of the cosmic large-scale structures, such as galaxies, through large-area homogeneous scans of the sky. The historical origin of such search strategies is inherently related to the theoretical formulation of cosmology. Without any detailed knowledge about the actual spatial distribution of cosmic matter, theorists resorted to predicting the mean value of statistical summaries, constituting ensemble averages over different realizations of the unknown cosmic matter distribution. Until today, most cosmological analyses are performed on the basis of two-point statistics \citep{tegmark2004threedimensional, percival2007shape, porredon2021dark}, although there are ongoing efforts to go beyond two- or three-point analyses. The initial need to use statistical summaries was twofold. First, the calculation of detailed matter field realizations by evaluating gravitational structure growth via simulations was costly and complex. Therefore, cosmological perturbation theory provided an efficient alternative to make predictions on the average statistical properties of the cosmic structures \citep{buchert1994testing, bouchet1995perturbative}. Secondly, little was known about the particular realization of the cosmic matter distribution of our Universe. The best way forward was to test statements that would be true on average using the quantitative diagnostics provided by statistical summaries.

\begin{figure*}[t]
	\centering
		{\includegraphics[width=\hsize,clip=true]{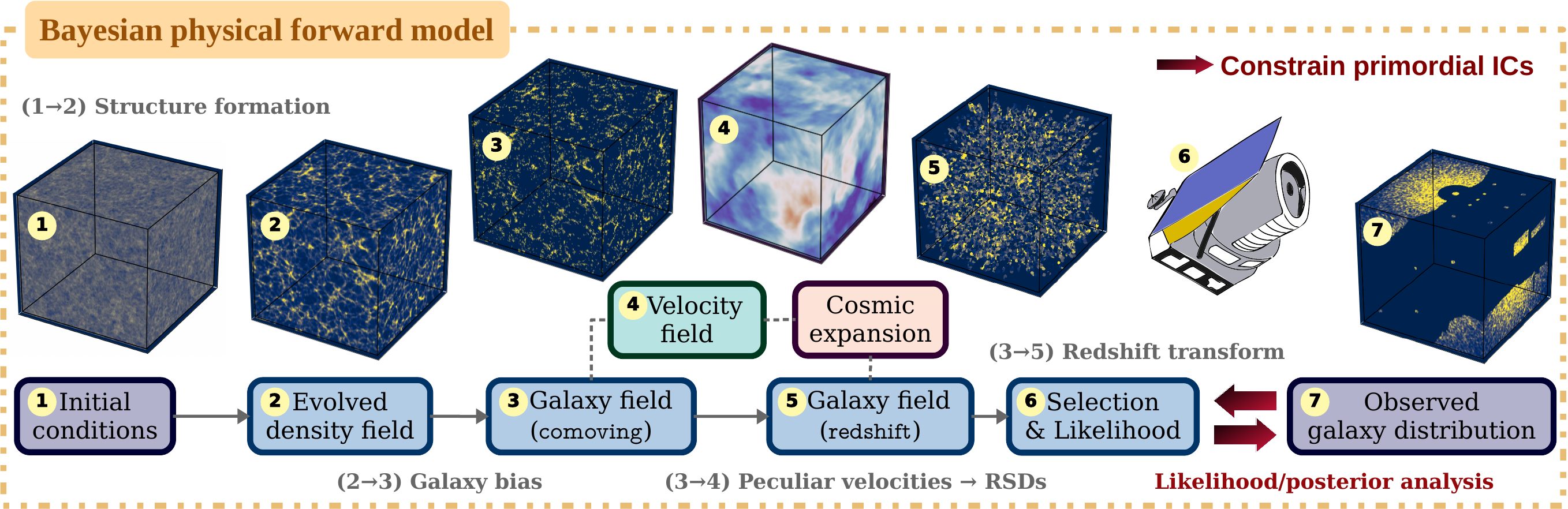}}
	\caption{Schematic of the Bayesian physical forward modeling framework of \borg. \borg solves a large-scale Bayesian inverse problem by fitting a dynamical structure formation model to galaxy observations and subsequently inferring the primordial initial conditions (ICs) which lead to the formation of the presently observed cosmic structures via gravitational evolution. The \textsc{borg} forward modeling approach marginalizes naturally over unknown galaxy bias, and accounts for all relevant physical effects, such as the redshift space distortions (RSDs) resulting from the peculiar velocities of galaxies, and instrumental selection effects.}
	\label{fig:borg_schematic}
\end{figure*}

The above situation has witnessed a dramatic improvement in the era of modern cosmology. The abundance of cosmological observations and the availability of computational resources have now contributed significantly in constructing a much more refined picture of the actual spatial configuration of the cosmic structures beyond characterizations via statistical summaries. In particular, we have proposed the Bayesian physical forward modeling framework \citep{jasche2013bayesian,lavaux2016unmasking, jasche2019physical} as a novel method enabling us to reconstruct the full 3D density field underlying observed galaxies in surveys with high fidelity. This machinery performs a causal inference that is informed by physics of the cosmic large-scale structures, their initial conditions and dynamical evolution. By characterizing the full 3D density field, our inference framework exploits the information on the phase distribution and higher-order statistics in data, which is inaccessible to traditional approaches. Although analyses relying on the use of the information content contained in the phase distribution of the large-scale structures have been proposed since a long time \citep[see, for e.g.,][]{chiang2000phase, byun2020constraining}, it was not propagated in the same manner as in the aforementioned physical forward modeling machinery. Therefore, the inference carried out in previous studies by \citet{jasche2013bayesian,lavaux2016unmasking, jasche2019physical} transcends the limitations of contemporary analyses based on statistical summaries in order to yield detailed characterizations of individual 3D structures. Presently, apart from our framework, several research groups are developing the technology to perform full 3D characterizations of the cosmic structures probed by galaxy surveys \citep{wang2014ELUCID, modi2018cosmological, kitaura2020cosmic, porqueres2020hierarchical, porqueres2020shear}.

So far, most of the work in the literature is focused on constraining cosmological physics from existing data sets, such as CMB \citep[see, for e.g.,][and references therein]{planck2018cosmo} and large-scale structure \citep[e.g.][]{einasto2010wavelet, byun2020constraining, porredon2021dark} observations. Here, we want to go beyond the standard task of constraining models to answering the question of how to optimally acquire future data that will be most informative to update our cosmological knowledge and to study physics. In doing so, we use existing information on cosmic structures to identify regions in the sky that promise the highest discovery potential as quantified by the Fisher information. Given the advent of the new data analysis technologies, we think that now it is especially timely to start a debate on whether the standard cosmological search strategy should be revised.

We note that some past studies aimed at understanding if certain regions of the sky are particularly informative about cosmology \citep[e.g.][]{mukherjee2018making}, or if one could use such information to design better survey geometries in the future \citep{bassett2005optimizing}. The method proposed here goes beyond these studies by using the full forward model of the 3D matter field for quantifying the information content, weighted by the posterior of plausible realizations of the large-scale structures in the Universe.

In summary, we would like to address the following questions: Is it true that a homogeneous scan of the sky is the optimal search strategy to test fundamental physics with cosmological surveys? Are there some particular regions in the sky that are more informative than others with respect to specific research questions? Is it possible to answer some research questions faster with small, cheap and targeted searches? These questions are relevant, as they in turn raise questions on the optimal use of scientific resources. For example, two upcoming large-scale cosmological surveys, the Vera Rubin Observatory \citep{ivezic2019lsst} and Euclid \citep{euclid2016missiondesign}, constitute a total financial investment of over a billion dollars. As such, to optimize the scientific returns of these missions, we must ensure that the limited survey resources are adequately managed and optimally utilized.

With the ideas outlined in this work, we wish to trigger a discussion in the cosmological community on whether scientific progress in terms of information gain on fundamental physics can be sped up by using more refined cosmological searches, akin to active learning strategies. We also hope to initiate new technological developments for targeted searches of cosmological physics. To advance in this direction, here, we address the following question: ``Which parts of the Universe should we observe to optimally retrieve information about a research question of interest?'' Specifically, in this work, we are interested in identifying the regions in the Universe that are most relevant to obtain new information about the cosmological parameters underlying our standard model of cosmology. But our approach is equally applicable to other research questions of interest.

We present, for the first time, maps detailing the expected information gain provided by cosmic structures of the nearby Universe on cosmological parameters, based on reconstructions of the 2M++ galaxy catalog \citep{LH11}. Our proposed methodology is, nevertheless, not limited to the nearby Universe and is applicable to any existing cosmological data sets.

\begin{figure*}[t]
	\centering
    \subfloat{\includegraphics[width=0.31\hsize]{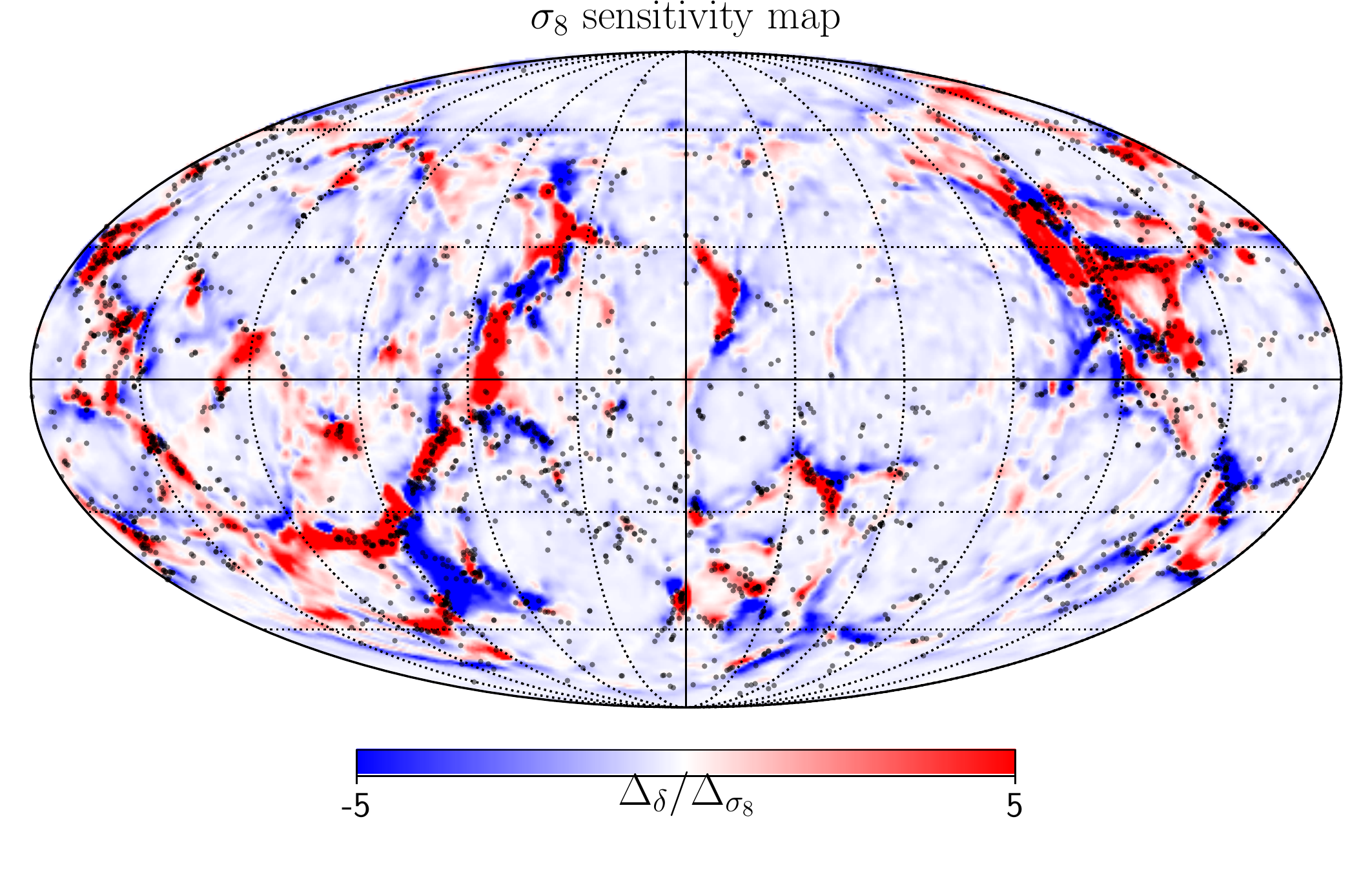}}
    \! \! \! \!
    \subfloat{\includegraphics[width=0.31\hsize]{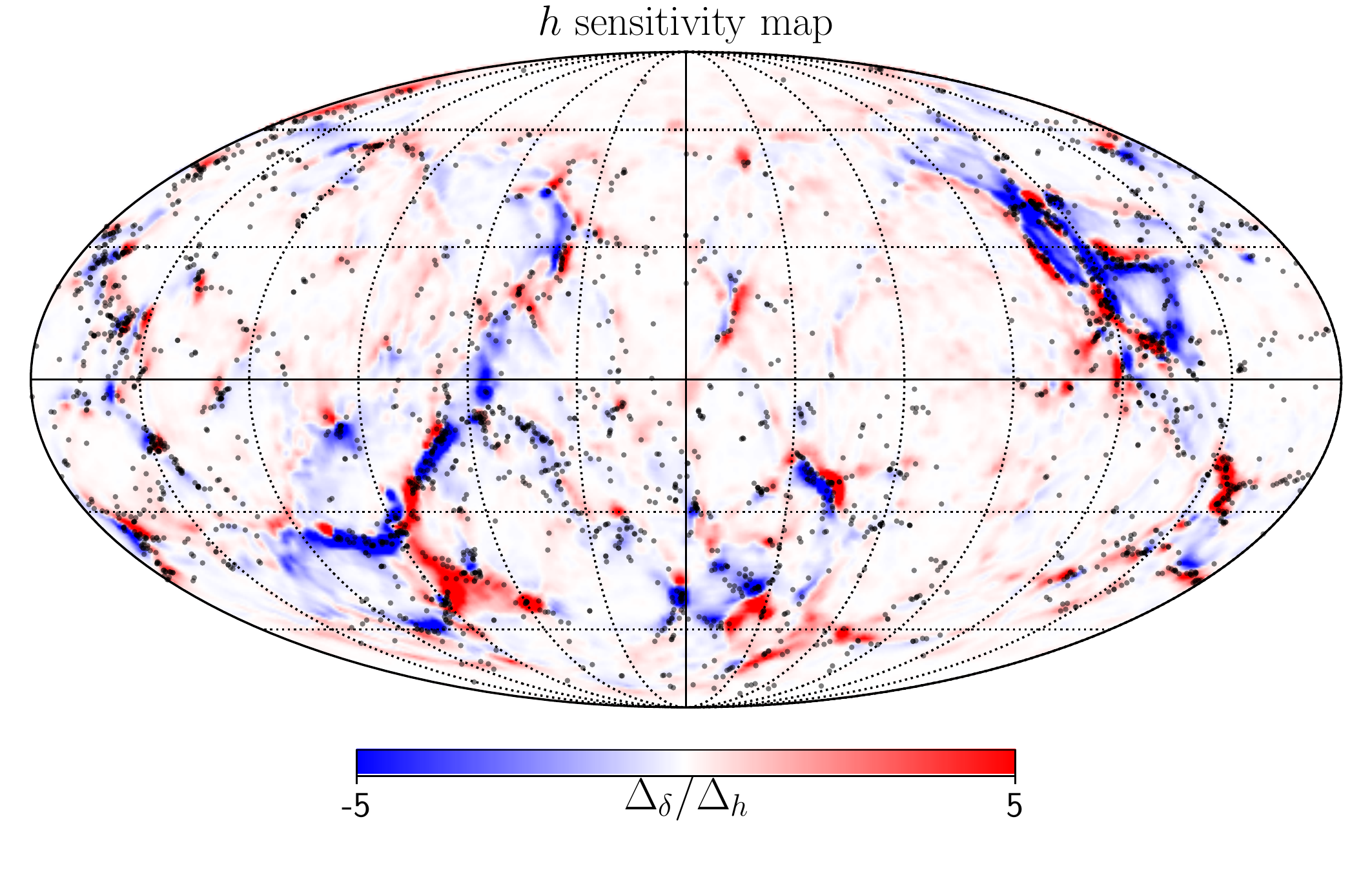}}
    \! \! \! \!
    \subfloat{\includegraphics[width=0.31\hsize]{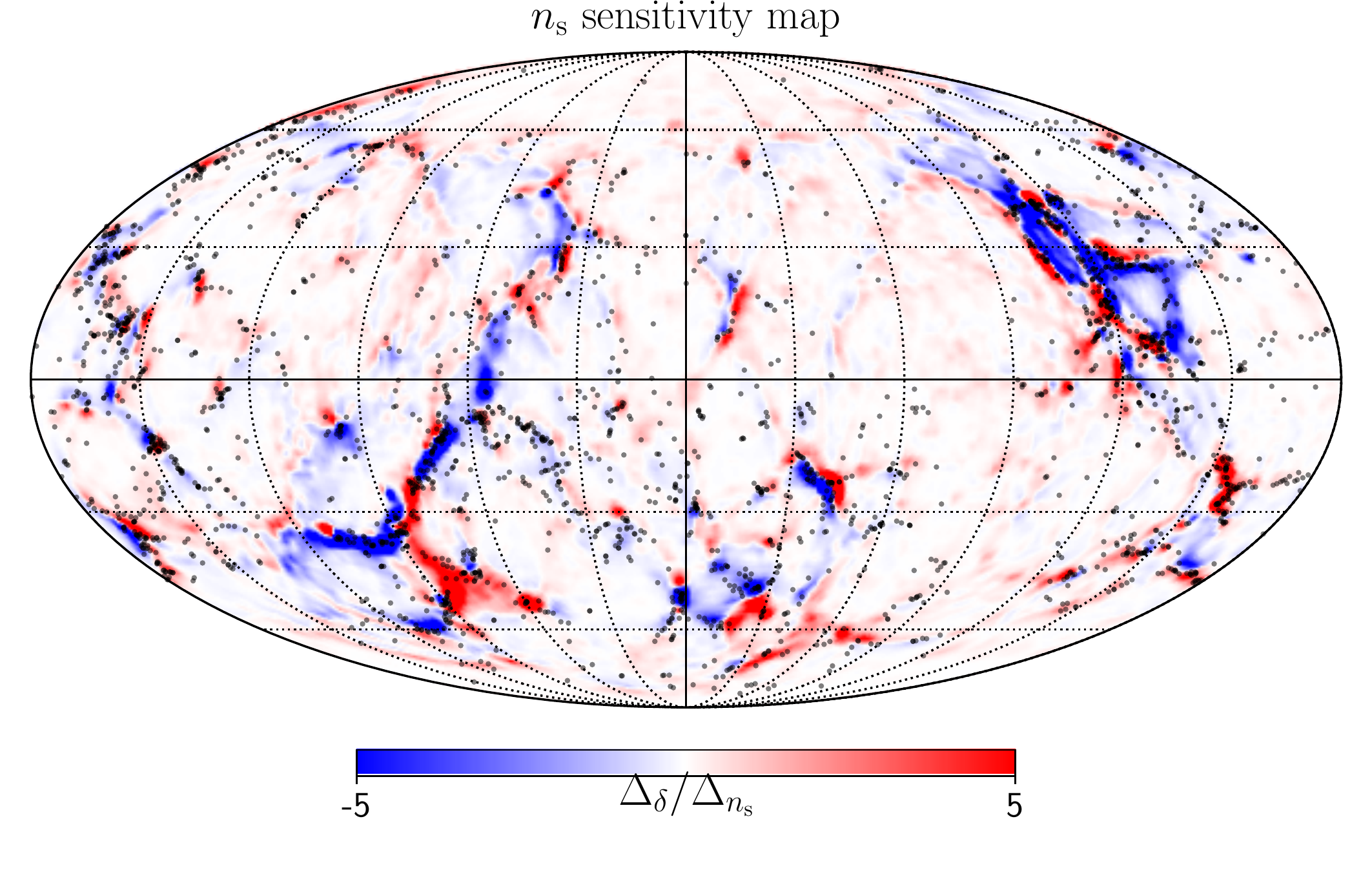}}
    \! \! \! \!
    \subfloat{\includegraphics[width=0.31\hsize]{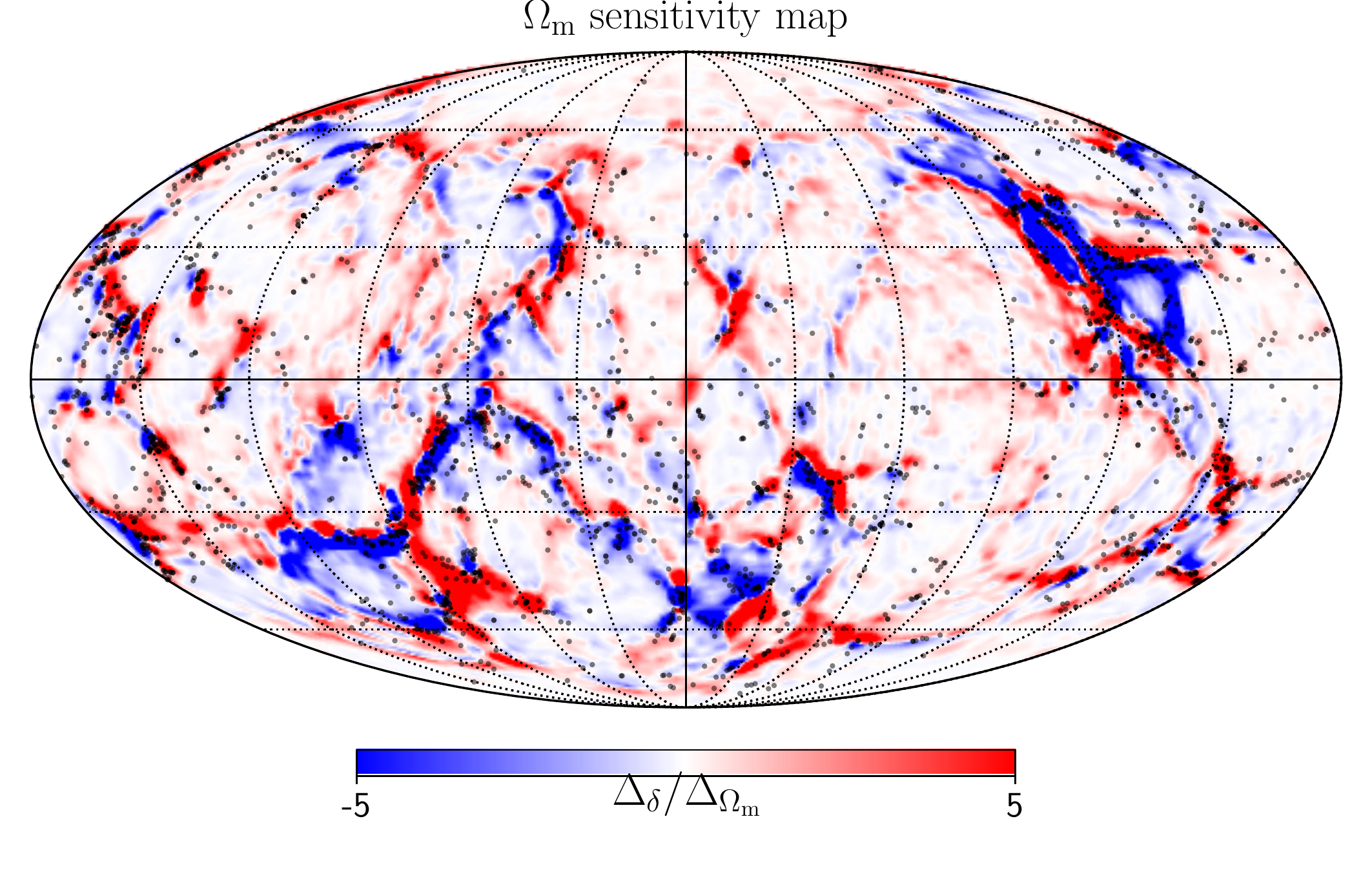}}
    \! \! \! \!
    \subfloat{\includegraphics[width=0.31\hsize]{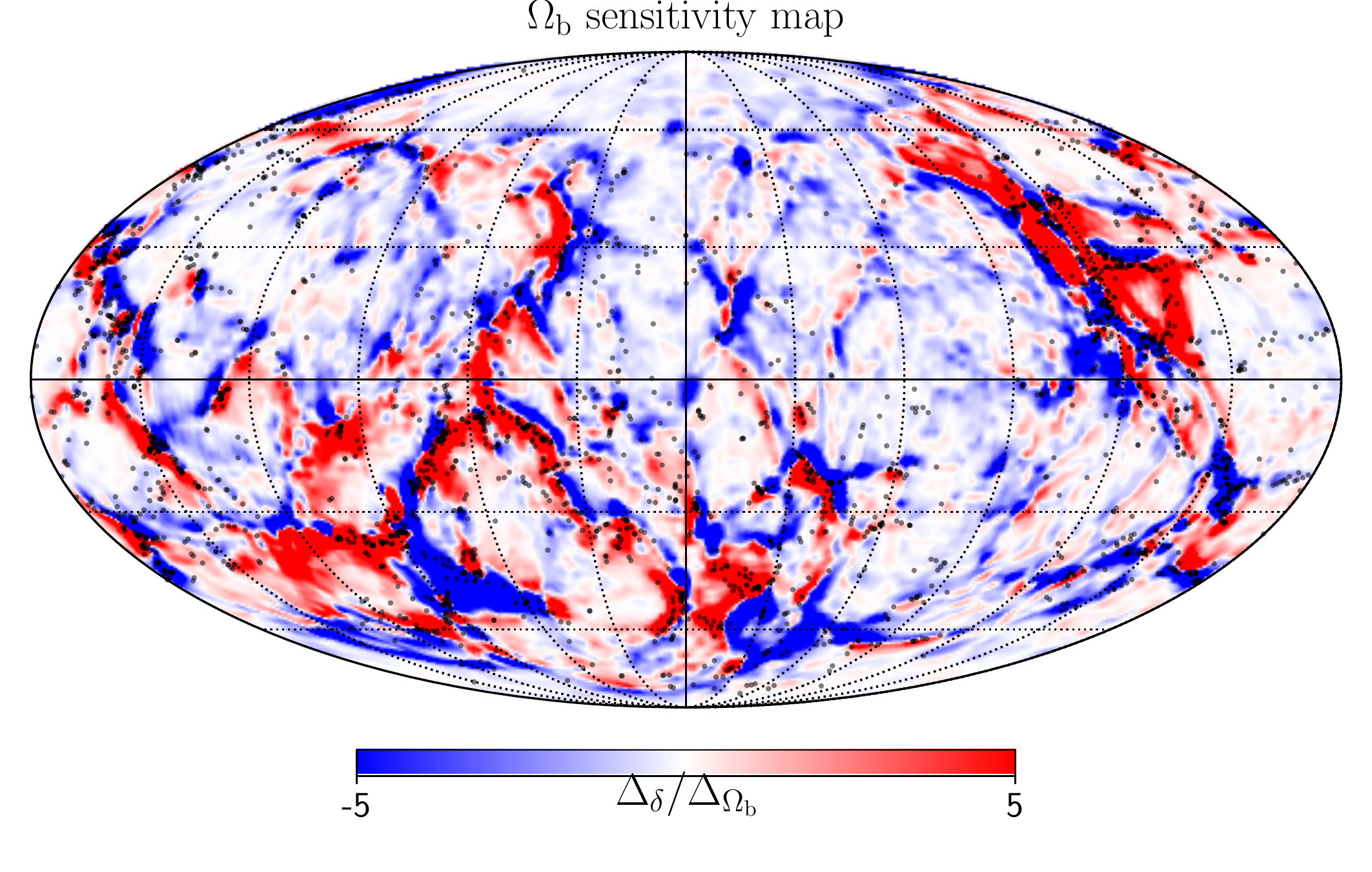}}
    \! \! \! \!
    \subfloat{\includegraphics[width=0.31\hsize]{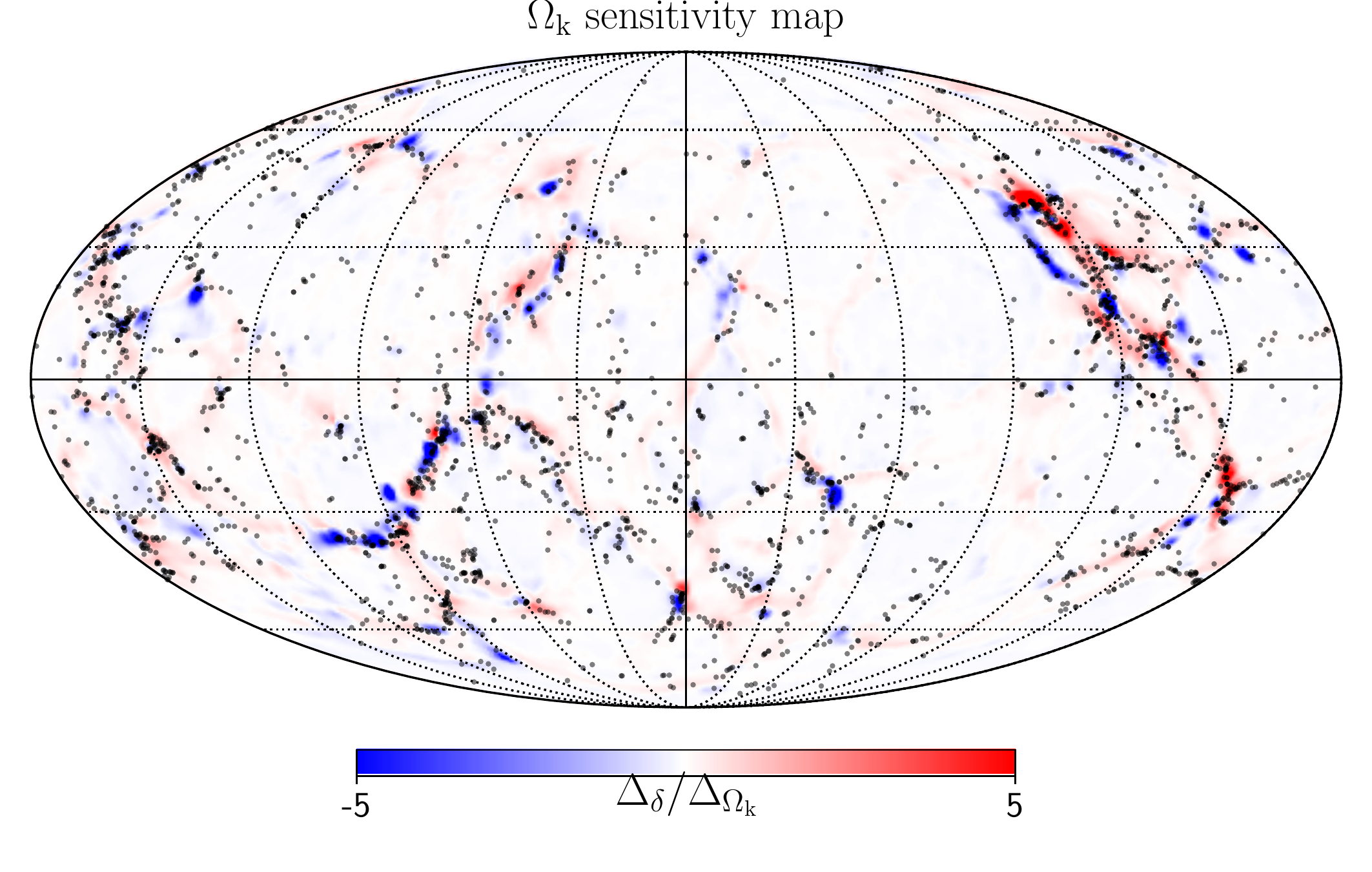}}
	\caption{Components of the gradient of the matter density field with respect to the cosmological parameters for a spherical slice of thickness $\sim 2.65$ \Mpch at a comoving distance of 100\Mpch from the observer, with the corresponding galaxies from the 2M++ catalog denoted via black dots. Visually, we find that the baryon density $\Omega_{\mathrm{b}}$ has the largest influence on the spatial distribution of the cosmic structures, with the regions surrounding the filamentary galaxy distribution being particularly sensitive to changes in the baryon density. Conversely, the least significant response emanates from the cosmic curvature $\Omega_{\mathrm{k}}$, with only the vicinity of the dense galaxy clusters reacting to changes in the geometry of the Universe.}
	\label{fig:dmaps_PM}
\end{figure*}
%--------------------------------------------------------------------

\section{Results}
\label{results}

In our quest for mapping out the cosmologically sensitive large-scale structures in the sky, we adopt the Fisher information \citep{fisher1925theory} methodology to provide a quantitative measure of the information encoded by the cosmological parameters on the observed large-scale structures. Although computing the Fisher information in this context is a highly non-trivial task, this ambitious undertaking is rendered feasible by employing a physical forward modeling machinery, such as our \borg (Bayesian Origin Reconstruction from Galaxies) algorithm \citep{jasche2013bayesian, jasche2019physical}. \borg constitutes a causal model of structure formation with a fully non-linear treatment of the dark matter clustering, allowing for the connection between the cosmic initial conditions and the observed galaxy distribution. A schematic view of the distinct components of the \borg forward model is illustrated in Fig.~\ref{fig:borg_schematic}. In our study, we marginalize over the ensemble of plausible realizations of the 3D primordial matter fluctuations of the very early Universe, as inferred by \borg via a hierarchical Bayesian statistical inference framework, from the 2M++ galaxy catalog \citep{LH11} that traces the matter distribution of the nearby Universe. We refer the reader to the appendices for more detailed information pertaining to the \borg algorithm, 2M++ galaxy catalog, mathematical formalism underlying the Fisher information map and numerical implementation thereof. We stress that our approach for deriving the Fisher information map, based on the constrained realizations of cosmic structures conditional on galaxy observations, bears a stark contrast to the standard Fisher analyses to obtain forecasts on cosmological constraints from forthcoming galaxy surveys.

The availability of such inferred primordial matter fluctuations enables us to test the causal sensitivity of cosmic structures when forward modeled with perturbed cosmological parameters. To illustrate this possibility, in Fig.~\ref{fig:dmaps_PM}, we illustrate the response of cosmic structures in the Universe to perturbative changes in six parameters of the so-called concordance $\Lambda$CDM cosmological model, namely the matter density $\Omega_{\mathrm{m}}$, baryon density $\Omega_{\mathrm{b}}$, cosmic curvature $\Omega_{\mathrm{k}}$, Hubble constant $h$, amplitude of matter fluctuations $\sigma_8$ and scalar spectral index $n_{\mathrm{s}}$ of the primordial power spectrum. This entails computing the gradient of the 3D galaxy field with respect to the cosmological parameters. This gradient quantifies the sensitivity of the galaxy distribution to changes in the cosmological parameters. As such, the gradient is described by a 6D vector for each volume element in the 3D grid. We compute the gradient using finite difference by executing \borg forward model evaluations on the ensemble of data constrained realizations of the cosmic initial conditions from the \borg 2M++ analysis \citep{jasche2019physical}, while varying the cosmological parameters about their corresponding fiducial values, as given by the latest best-fit values from the Planck Collaboration \citep{planck2018cosmo}.

\begin{figure*}[t]
	\centering
    \subfloat{\includegraphics[width=0.72\hsize, clip=true]{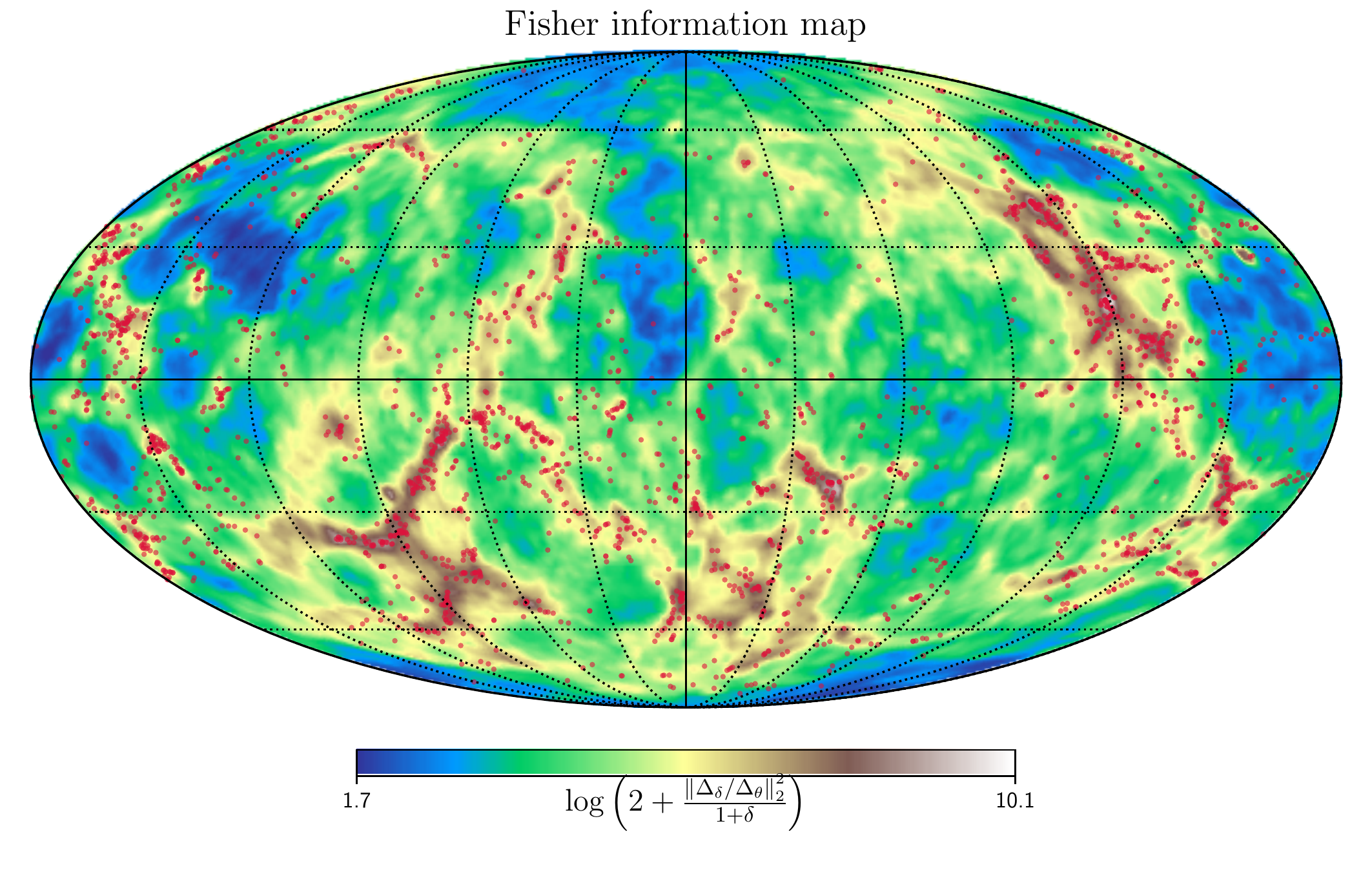}}
	\caption{Fisher information map for the same spherical slice as in Fig.~\ref{fig:dmaps_PM}. The observed galaxy distribution from the 2M++ catalog, lying in the corresponding spherical shell centered around the observer, is represented by the red dots. We find that the regions in the vicinity of the massive cosmic structures, as traced by the galaxy distribution, are the most informative according to the Fisher information map. These regions correspond to the regime of gravitational infall of the galaxy clusters.}
	\label{fig:fisher_maps_PM}
\end{figure*}

In Fig.~\ref{fig:dmaps_PM}, we visualize the individual gradient components of the six cosmological parameters by representing them as {\it cosmological sensitivity maps}. To this end, we consider a spherical slice of thickness $\sim 2.65$ \Mpch at a comoving distance of 100\Mpch from the observer that is projected onto a \textsc{healpix} map \citep{gorski2005healpix}. The observed galaxies from the 2M++ catalog lying in the projected spherical slice are also indicated. From a visual comparison, we find that the density of baryonic matter, as characterized by the $\Omega_{\mathrm{b}}$ parameter, induces the most significant response in the observed distribution of cosmic structures, with the $\Omega_{\mathrm{m}}$ and $\sigma_8$ parameters also having a notable influence. The astrophysical regions surrounding the filamentary patterns traced by the galaxies respond effectively to changes in the latter cosmological parameters. In contrast, the cosmic matter distribution displays a relatively minimal sensitivity to the $\Omega_{\mathrm{k}}$ parameter, with the most noticeable effect of the cosmic curvature on the large-scale structures being on the densest regions, such as the galaxy clusters. The underdense regions of the matter distribution, the so-called cosmic voids, are most impacted by changes in the $\Omega_{\mathrm{b}}$ and $\sigma_8$ parameters. We clarify that our structure formation model does not account for baryonic physics, so that $\Omega_{\mathrm{b}}$ only modifies the amplitude of growth of matter perturbations and the shape of the linear power spectrum. Nevertheless, it is straightforward to repeat our analysis with hydrodynamical simulations to properly account for the presence of baryons. Given the significantly higher computational cost of such simulations, one possible option would be to use physics emulators, recently being developed on a large scale \citep{villaescusa2020camels}.

\begin{figure}
	\centering
		{\includegraphics[width=\hsize,clip=true]{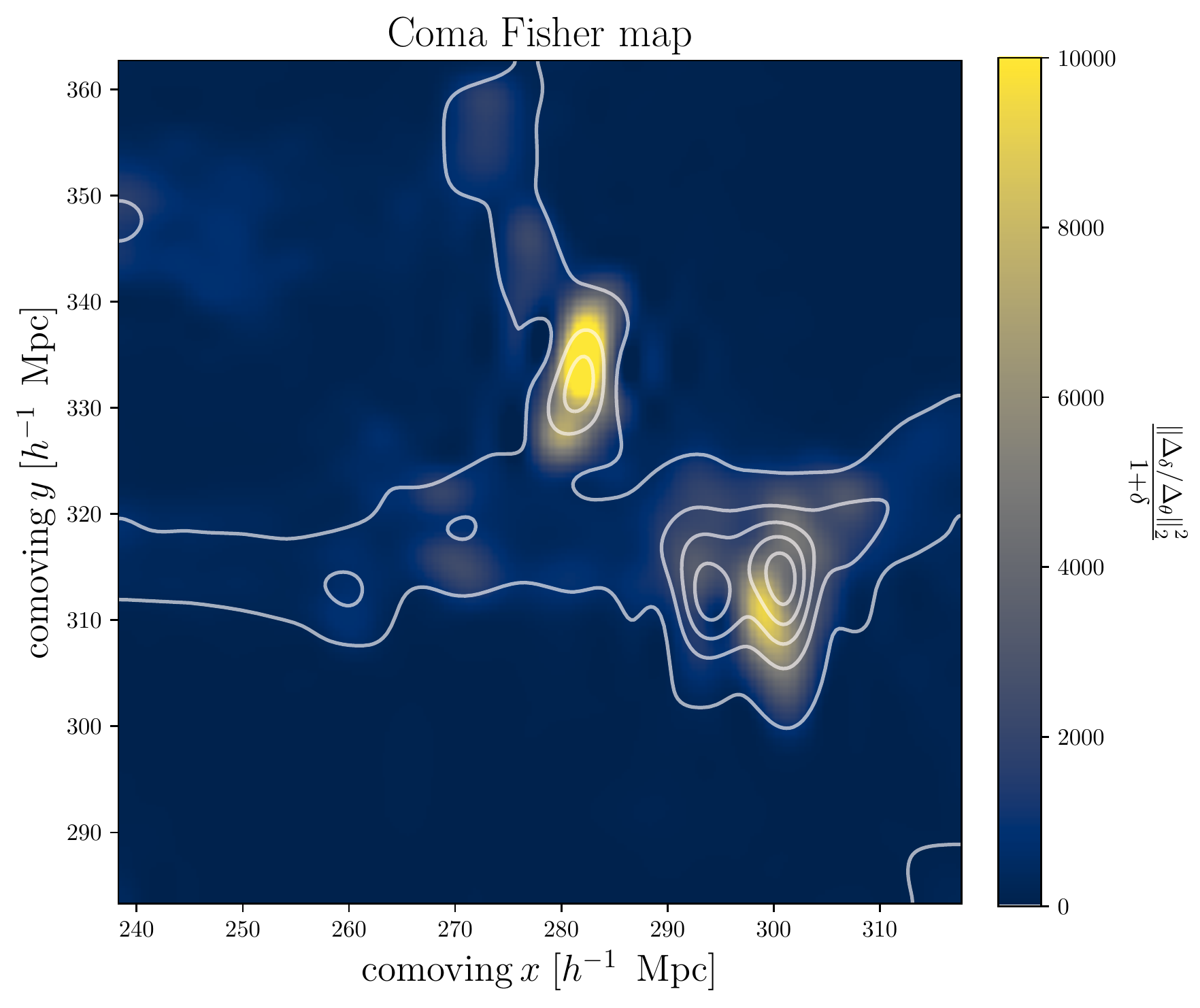}}
	\caption{Fisher information map of the Coma cluster, with its corresponding mass density overlaid as a contour, for the central slice of thickness $\sim 5$\Mpch through a 3D region centered on the cluster that extends over $40$\Mpch. According to the Fisher information map, the central region containing the core of the cluster encodes fairly limited information gain, whilst the peripheral regions of the filaments and cluster core, where mass accretion occurs via gravitational infall, constitute the largest amount of information gain.}
	\label{fig:coma_fisher}
\end{figure}

In addition to displaying the relative strength of the causal response of the cosmic structures to changes in our underlying cosmological model, the array of sensitivity maps in Fig.~\ref{fig:dmaps_PM} reveals some interesting features. The $\sigma_8$ map, for example, illustrates how the clustering of matter occurs at the expense of the neighbouring regions, while the $h$ map shows that a modified Hubble flow yields different clustering features. Moreover, the $\Omega_{\mathrm{m}}$ and $\Omega_{\mathrm{b}}$ maps display distinct anti correlated signatures. This may be attributed to the fact that the sound horizon distance scale due to baryon acoustic oscillations (BAOs) increases with both $\Omega_{\mathrm{m}}$ and $\Omega_{\mathrm{b}}$, with the effect induced by $\Omega_{\mathrm{b}}$ being much stronger, such that a fixed BAO scale encoded in the data yields this anti correlation between these two maps. Similarly, the $\Omega_{\mathrm{m}}$ and $\sigma_8$ maps depict a striking anti correlation. This can be naturally explained on the basis that the primary constraining power emanates from the combination of the growth rate $f$ of cosmological perturbations and $\sigma_8$, where $f \sigma_8 \approx \sigma_8 \Omega_{\mathrm{m}}^{0.55}$ for the $\Lambda$CDM model.

A crucial ingredient in the Fisher information formalism derived in our study, as described by Eq.~\eqref{eqn: finite_difference_maps_fisher}, lies in computing the above gradient. The desired 3D Fisher information field, obtained using Eq.~\eqref{eqn: finite_difference_maps_fisher}, can be represented as a {\it Fisher information map} in the same way as the cosmological sensitivity maps. This Fisher information map represents the combined information gain on all six cosmological parameters considered in this study. The Fisher information map, for a spherical slice of thickness $\sim 2.65$\Mpch at a distance of 100\Mpch from the observer, is displayed in Fig.~\ref{fig:fisher_maps_PM}, along with the observed galaxy distribution in this particular slice. The Fisher information map indicates that the regions with the highest information gain are those in the vicinity of the filamentary distribution traced by galaxies. One plausible interpretation is that the regions of gravitational infall, where there is the accumulation of matter, are the most informative about our cosmological model. This is particularly interesting as these infalling regions are, as yet, relatively obscure and must be properly understood.

The \borg analysis of the 2M++ galaxy catalog showcased the capacity of the physical forward modeling machinery to resolve the key features of prominent cosmic structures in the present Universe \citep{jasche2019physical}. The inferred mass profile of the Coma cluster, in particular, was found to be in remarkable agreement with state-of-the-art weak lensing measurements. Therefore, it is also interesting to study the characteristic features underlying the source of cosmological information for this well-known cluster. Fig.~\ref{fig:coma_fisher} displays the Fisher information map of the Coma cluster and its corresponding mass density. We note that the figure depicts the central slice of thickness $\sim 5$\Mpch through a 3D patch extending over $40$\Mpch. The features present in the Fisher map of the Coma cluster are in accordance with those from the sky projected Fisher map from Fig.~\ref{fig:fisher_maps_PM}, thereby supporting the interpretation that the information that can be gleaned from the regions surrounding the filaments, i.e. the gravitationally infalling regions of the cluster, are the most significant. Recent studies, relying on conceptually distinct methodologies, also indicate that accretion filaments and the surroundings of voids are highly sensitive to predictions of dark energy \citep{leclercq2016comparing} and gravity \citep{lam2012testing} models. In contrast, the central core of the cluster provides a relatively lower information gain, with the regions devoid of matter displaying unsubstantial information. A key point worth stressing here is that our relatively simple physics informed algorithm was capable of pinning down potential regions of cosmological interest in the sky.

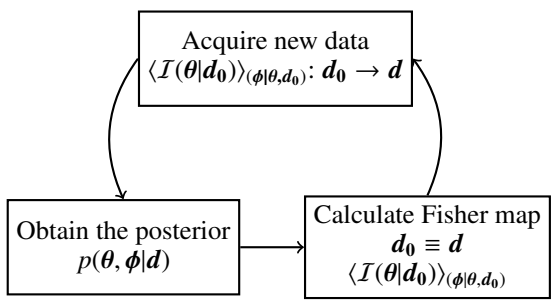
\begin{figure}
\centering
\begin{tikzpicture}[thick, node distance=4cm]
    \node[targetsearchstep] (A) at (2,2.5){Acquire new data\\$\langle\mathcal{I}(\mvec{\theta}|\mvec{d_0})\rangle_{(\mvec{\phi|\mvec{\theta},\mvec{d_0}})}$: $\mvec{d_0} \rightarrow \mvec{d}$};
    \node[targetsearchstep] (B) at (0,0){Obtain the posterior\\$p(\mvec{\theta}, \mvec{\phi} | \mvec{d})$};
    \node[targetsearchstep] (C) at (4,0){Calculate Fisher map\\
    $\mvec{d_0} \equiv \mvec{d}$\\
    $\langle\mathcal{I}(\mvec{\theta}|\mvec{d_0})\rangle_{(\mvec{\phi}|\mvec{\theta},\mvec{d_0})}$};
%------------------
    \draw[->] (A.west) to[bend right] (B.north) ;
    
    \draw[->] (B.east) -- (C.west);
    
    \draw[->] (C.north) to[bend right](A.east);
\end{tikzpicture}
\caption
{Flowchart of the targeted search approach. The core idea is to use the existing knowledge as quantified by the posterior $p(\mvec{\theta},\mvec{\phi}|\mvec{d})$ in order to calculate the Fisher map marginalized over this posterior $\langle\mathcal{I}(\mvec{\theta}|\mvec{d_0})\rangle_{(\mvec{\phi}|\mvec{\theta},\mvec{d_0})}$, keeping the observables of interest. Then, this Fisher map will provide us with regions of the sky with highest information gain potential for acquiring new data optimal for testing our model predictions, $\langle\mathcal{I}(\mvec{\theta}|\mvec{d}_0)\rangle_{(\mvec{\phi|\mvec{\theta},\mvec{d_0}})}$: $\mvec{d_0} \rightarrow \mvec{d}$. This data will in turn be used to update our knowledge about the model through updating the posterior and the procedure would repeat until the information content is fully depleted.}
\label{fig: target_search_flowchart}
\end{figure}

Now, one can use the maps displayed in Figs.~\ref{fig:fisher_maps_PM} and \ref{fig:coma_fisher} as a guide towards where to look in order to optimally collect data for testing our physical model of structure formation. For the particular case demonstrated here, the data consists of galaxy counts and the corresponding information gain predictions are for the galaxy clustering data. The idea behind the targeted search approach is to recursively search for galaxies in the high information gain regions, as quantified by the Fisher map, then use this newly acquired data to constrain the model parameters and repeat the procedure. This is schematically depicted in Fig.~\ref{fig: target_search_flowchart}. The justification pertaining to why this search strategy is optimal is further elaborated in Appendix~\ref{app: appendixD}. It should be understood that this approach can be similarly extended to other observables for constructing the corresponding information gain maps.

\section{Discussion}
\label{discussion}

The results presented in this work demonstrate the feasibility of machine aided targeted searches for cosmological physics signals. These have become feasible through recent developments of physics informed causal inference frameworks to study the 3D cosmic large-scale structures, their origin and evolution with time. We have used 3D initial conditions, inferred with the \borg{} algorithm, and a physics simulator of cosmological structure formation to chart the response of observed structures in the Universe and their corresponding information gain with respect to cosmological parameters. Our results are the first of their kind and elucidate the inhomogeneous distribution of cosmological information in the Universe. This study paves a new way forward to perform efficient targeted searches for the fundamental physics of the Universe, where search strategies are progressively refined with new cosmological data sets within an active learning framework.

We have further illuminated the response of the cosmic large-scale structures with respect to individual cosmological parameters. These results suggest that different features of the cosmic structures respond differently to perturbations in the physics determined by the cosmological parameters. For instance, we find that the vicinity of the filamentary cosmic structures, corresponding to gravitationally infalling regions, are highly sensitive to changes in the baryon density, with the cosmic curvature impacting only the surroundings of the dense galaxy clusters. Cosmic voids are primarily affected by the amplitude of matter fluctuations and the baryon density.  Our results demonstrate the value of going beyond state-of-the-art analyses of the cosmic large-scale structures that are limited to summary statistics and ignore this richness of the 3D cosmic structures. Even though we consider one particular observable, namely galaxy counts, in this study, our proposed framework can be seamlessly applied to other tracers, such as Lyman-$\alpha$ forest \citep{porqueres2019inferring, porqueres2020hierarchical}, which would bear complementary information.

Our findings further suggest that optimal targeted searches for research questions have become feasible. Given a specific research question, our approach enables us to propose targets for optimal information retrieval. This raises the question if traditional survey strategies should be revised and if significant scientific progress could also be driven by targeted searches with smaller, cheaper and faster instrumentation. We hope that our contribution will trigger a discussion and new technological advances in the field. 

While we studied the influence of cosmological parameters on the Universe in this work, our proof of concept reveals the potentially far-reaching and groundbreaking implications when considering the physical effects induced by modified gravity \citep{koyama2016cosmological}, dynamical dark energy \citep{zhao2017dynamical} and exotic dark matter models, such as self-interacting \citep{carlson1992SIDM} and fuzzy dark matter \citep{hu2000fuzzy}, or massive neutrinos \citep{paco2014neutrino}, on the cosmic large-scale structures. Once we identify the astrophysical region(s) of interest, based on the Fisher information map, for a particular model, we may subsequently proceed by computing accurate theoretical predictions for the spatial distribution of matter and luminous tracers for the given model via cosmological $N$-body or hydrodynamical simulations with extremely high resolution, thereby resulting in highly detailed physical features. The final step would then entail observing the relevant region(s) and comparing the theoretical predictions with the galaxy observations in a likelihood or posterior analysis. In essence, we are proposing a novel way of doing science that optimizes scarce resources for maximal scientific returns via an efficient observational strategy.

{\small {\it Acknowledgements.} We express our appreciation to Fabian Schmidt, Benjamin Wandelt, Eleni Tsaprazi, Natalia Porqueres, Minh Nguyen, Radek Wojtak, Florent Leclercq and Harry Desmond for their remarks on our manuscript. AK acknowledges support from the Starting Grant (ERC-2015-STG 678652) ``GrInflaGal'' of the European Research Council at MPA. JJ acknowledges support by the Swedish Research Council (VR) under the project 2020-05143 -- "Deciphering the Dynamics of Cosmic Structure".   DKR is a DARK fellow supported by a Semper Ardens grant from the Carlsberg Foundation (reference CF15-0384). This work was supported by the ANR BIG4 project, grant ANR-16-CE23-0002 of the French Agence Nationale de la Recherche. This work has made use of the Horizon Cluster hosted by Institut d'Astrophysique de Paris. This work has been done within the activities of the Domaine d'Int\'er\^et Majeur (DIM) Astrophysique et Conditions d'Apparition de la Vie (ACAV), and received financial support from R\'egion Ile-de-France. We thank Cepheid Studio (\url{https://www.cepheidstudio.com}) for providing us with the telescope illustration used in Fig.~\ref{fig:borg_schematic}. This work is done within the Aquila Consortium (\url{https://aquila-consortium.org}).}

{\small {\it Author contributions.} \textbf{AK}: led the project; methodology; software; obtained, validated, and interpreted results; writing - editing. \textbf{JJ}: project conceptualization; methodology; validation \& interpretation; supervision; writing - editing. \textbf{DKR}: methodology; visualization; validation \& interpretation, writing - editing. \textbf{GL}: methodology; validation \& interpretation; supervision; provided resources; funding acquisition, editing.}

% WARNING
%-------------------------------------------------------------------
% Please note that we have included the references to the file aa.dem in
% order to compile it, but we ask you to:
%
% - use BibTeX with the regular commands:
   \bibliographystyle{aa} % style aa.bst
   \bibliography{smap} % your references Yourfile.bib

\providecommand{\noopsort}[1]{}\providecommand{\singleletter}[1]{#1}%
\begin{thebibliography}{53}
\expandafter\ifx\csname natexlab\endcsname\relax\def\natexlab#1{#1}\fi

\bibitem[{{Abazajian} {et~al.}(2009){Abazajian}, {Adelman-McCarthy},
  {Ag{\"u}eros}, {Allam}, {Allende Prieto}, {An}, {Anderson}, {Anderson},
  {Annis}, {Bahcall}, \& et~al.}]{SDSS7}
{Abazajian}, K.~N., {Adelman-McCarthy}, J.~K., {Ag{\"u}eros}, M.~A., {et~al.}
  2009, ApJS, 182, 543

\bibitem[{{Bartlett} {et~al.}(2021){Bartlett}, {Desmond}, \&
  {Ferreira}}]{bartlett2021constraints}
{Bartlett}, D.~J., {Desmond}, H., \& {Ferreira}, P.~G. 2021, \prd, 103, 023523

\bibitem[{Bassett(2005)}]{bassett2005optimizing}
Bassett, B.~A. 2005, Physical Review D, 71, 083517

\bibitem[{{Bouchet} {et~al.}(1995){Bouchet}, {Colombi}, {Hivon}, \&
  {Juszkiewicz}}]{bouchet1995perturbative}
{Bouchet}, F.~R., {Colombi}, S., {Hivon}, E., \& {Juszkiewicz}, R. 1995, A\&A,
  296, 575

\bibitem[{{Buchert} {et~al.}(1994){Buchert}, {Melott}, \&
  {Weiss}}]{buchert1994testing}
{Buchert}, T., {Melott}, A.~L., \& {Weiss}, A.~G. 1994, A\&A, 288, 349

\bibitem[{{Byun} {et~al.}(2020){Byun}, {Franco}, {Howlett}, {Bonvin}, \&
  {Obreschkow}}]{byun2020constraining}
{Byun}, J., {Franco}, F.~O., {Howlett}, C., {Bonvin}, C., \& {Obreschkow}, D.
  2020, \mnras, 497, 1765

\bibitem[{{Carlson} {et~al.}(1992){Carlson}, {Machacek}, \&
  {Hall}}]{carlson1992SIDM}
{Carlson}, E.~D., {Machacek}, M.~E., \& {Hall}, L.~J. 1992, \apj, 398, 43

\bibitem[{{Charnock} {et~al.}(2020){Charnock}, {Lavaux}, {Wandelt}, {Sarma
  Boruah}, {Jasche}, \& {Hudson}}]{charnock2020neural}
{Charnock}, T., {Lavaux}, G., {Wandelt}, B.~D., {et~al.} 2020, MNRAS, 494, 50

\bibitem[{{Chiang} \& {Coles}(2000)}]{chiang2000phase}
{Chiang}, L.-Y. \& {Coles}, P. 2000, \mnras, 311, 809

\bibitem[{{Desmond} \& {Ferreira}(2020)}]{desmond2020galaxy}
{Desmond}, H. \& {Ferreira}, P.~G. 2020, \prd, 102, 104060

\bibitem[{{Desmond} {et~al.}(2018){Desmond}, {Ferreira}, {Lavaux}, \&
  {Jasche}}]{desmond2018fifth}
{Desmond}, H., {Ferreira}, P.~G., {Lavaux}, G., \& {Jasche}, J. 2018, \prd, 98,
  064015

\bibitem[{{Desmond} {et~al.}(2019){Desmond}, {Ferreira}, {Lavaux}, \&
  {Jasche}}]{desmond2019fifth}
{Desmond}, H., {Ferreira}, P.~G., {Lavaux}, G., \& {Jasche}, J. 2019, MNRAS,
  483, L64

\bibitem[{{Einasto} {et~al.}(2011){Einasto}, {H{\"u}tsi}, {Saar}, {Suhhonenko},
  {Liivam{\"a}gi}, {Einasto}, {M{\"u}ller}, {Starobinsky}, {Tago}, \&
  {Tempel}}]{einasto2010wavelet}
{Einasto}, J., {H{\"u}tsi}, G., {Saar}, E., {et~al.} 2011, \aap, 531, A75

\bibitem[{{Elsner} {et~al.}(2020){Elsner}, {Schmidt}, {Jasche}, {Lavaux}, \&
  {Nguyen}}]{elsner2020cosmology}
{Elsner}, F., {Schmidt}, F., {Jasche}, J., {Lavaux}, G., \& {Nguyen}, N.-M.
  2020, J. Cosmology Astropart. Phys., 2020, 029

\bibitem[{Fisher(1925)}]{fisher1925theory}
Fisher, R.~A. 1925, {Theory of statistical estimation} (Cambridge University
  Press)

\bibitem[{{G{\'o}rski} {et~al.}(2005){G{\'o}rski}, {Hivon}, {Banday},
  {Wandelt}, {Hansen}, {Reinecke}, \& {Bartelmann}}]{gorski2005healpix}
{G{\'o}rski}, K.~M., {Hivon}, E., {Banday}, A.~J., {et~al.} 2005, \apj, 622,
  759

\bibitem[{{Hu} {et~al.}(2000){Hu}, {Barkana}, \& {Gruzinov}}]{hu2000fuzzy}
{Hu}, W., {Barkana}, R., \& {Gruzinov}, A. 2000, \prl, 85, 1158

\bibitem[{{Huchra} {et~al.}(2012){Huchra}, {Macri}, {Masters}, {Jarrett},
  {Berlind}, {Calkins}, {Crook}, {Cutri}, {Erdo{\v g}du}, {Falco},
  {et~al.}}]{Huchra12}
{Huchra}, J.~P., {Macri}, L.~M., {Masters}, K.~L., {et~al.} 2012, ApJS, 199, 26

\bibitem[{{Ivezi{\'c}} {et~al.}(2019){Ivezi{\'c}}, {Kahn}, {Tyson}, {Abel},
  {Acosta}, {Allsman}, {Alonso}, {AlSayyad}, {Anderson}, {Andrew},
  {et~al.}}]{ivezic2019lsst}
{Ivezi{\'c}}, {\v{Z}}., {Kahn}, S.~M., {Tyson}, J.~A., {et~al.} 2019, \apj,
  873, 111

\bibitem[{{Jasche} \& {Kitaura}(2010)}]{jasche2010fast}
{Jasche}, J. \& {Kitaura}, F.~S. 2010, MNRAS, 407, 29

\bibitem[{{Jasche} \& {Lavaux}(2019)}]{jasche2019physical}
{Jasche}, J. \& {Lavaux}, G. 2019, A\&A, 625, A64

\bibitem[{{Jasche} \& {Wandelt}(2013)}]{jasche2013bayesian}
{Jasche}, J. \& {Wandelt}, B.~D. 2013, MNRAS, 432, 894

\bibitem[{{Jones} {et~al.}(2009){Jones}, {Read}, {Saunders}, {Colless},
  {Jarrett}, {Parker}, {Fairall}, {Mauch}, {Sadler}, {Watson},
  {et~al.}}]{Jones09}
{Jones}, D.~H., {Read}, M.~A., {Saunders}, W., {et~al.} 2009, MNRAS, 399, 683

\bibitem[{{Kitaura} {et~al.}(2021){Kitaura}, {Ata}, {Rodr{\'\i}guez-Torres},
  {Hern{\'a}ndez-S{\'a}nchez}, {Balaguera-Antol{\'\i}nez}, \&
  {Yepes}}]{kitaura2020cosmic}
{Kitaura}, F.-S., {Ata}, M., {Rodr{\'\i}guez-Torres}, S.~A., {et~al.} 2021,
  MNRAS, 502, 3456

\bibitem[{{Kodi Ramanah} {et~al.}(2019){Kodi Ramanah}, {Lavaux}, {Jasche}, \&
  {Wandelt}}]{DKR2018altair}
{Kodi Ramanah}, D., {Lavaux}, G., {Jasche}, J., \& {Wandelt}, B.~D. 2019, A\&A,
  621, A69

\bibitem[{{Koyama}(2016)}]{koyama2016cosmological}
{Koyama}, K. 2016, Reports on Progress in Physics, 79, 046902

\bibitem[{{Lam} {et~al.}(2012){Lam}, {Nishimichi}, {Schmidt}, \&
  {Takada}}]{lam2012testing}
{Lam}, T.~Y., {Nishimichi}, T., {Schmidt}, F., \& {Takada}, M. 2012, \prl, 109,
  051301

\bibitem[{{Lavaux} \& {Hudson}(2011)}]{LH11}
{Lavaux}, G. \& {Hudson}, M.~J. 2011, MNRAS, 416, 2840

\bibitem[{{Lavaux} \& {Jasche}(2016)}]{lavaux2016unmasking}
{Lavaux}, G. \& {Jasche}, J. 2016, MNRAS, 455, 3169

\bibitem[{{Leclercq} {et~al.}(2016){Leclercq}, {Lavaux}, {Jasche}, \&
  {Wandelt}}]{leclercq2016comparing}
{Leclercq}, F., {Lavaux}, G., {Jasche}, J., \& {Wandelt}, B. 2016, J. Cosmology
  Astropart. Phys., 2016, 027

\bibitem[{{Modi} {et~al.}(2018){Modi}, {Feng}, \&
  {Seljak}}]{modi2018cosmological}
{Modi}, C., {Feng}, Y., \& {Seljak}, U. 2018, J. Cosmology Astropart. Phys.,
  2018, 028

\bibitem[{{Mukherjee} {et~al.}(2021){Mukherjee}, {Lavaux}, {Bouchet}, {Jasche},
  {Wandelt}, {Nissanke}, {Leclercq}, \& {Hotokezaka}}]{mukherjee2019velocity}
{Mukherjee}, S., {Lavaux}, G., {Bouchet}, F.~R., {et~al.} 2021, A\&A, 646, A65

\bibitem[{{Mukherjee} \& {Wandelt}(2018)}]{mukherjee2018making}
{Mukherjee}, S. \& {Wandelt}, B.~D. 2018, \jcap, 2018, 042

\bibitem[{{Neyrinck} {et~al.}(2014){Neyrinck}, {Arag{\'o}n-Calvo}, {Jeong}, \&
  {Wang}}]{neyrinck2014halo}
{Neyrinck}, M.~C., {Arag{\'o}n-Calvo}, M.~A., {Jeong}, D., \& {Wang}, X. 2014,
  MNRAS, 441, 646

\bibitem[{{Nguyen} {et~al.}(2020){Nguyen}, {Jasche}, {Lavaux}, \&
  {Schmidt}}]{nguyen2020kSZ}
{Nguyen}, N.-M., {Jasche}, J., {Lavaux}, G., \& {Schmidt}, F. 2020, J.
  Cosmology Astropart. Phys., 2020, 011

\bibitem[{{Nguyen} {et~al.}(2021){Nguyen}, {Schmidt}, {Lavaux}, \&
  {Jasche}}]{nguyen2021impacts}
{Nguyen}, N.-M., {Schmidt}, F., {Lavaux}, G., \& {Jasche}, J. 2021, J.
  Cosmology Astropart. Phys., 2021, 058

\bibitem[{{Pardo} {et~al.}(2019){Pardo}, {Desmond}, \&
  {Ferreira}}]{pardo2019testing}
{Pardo}, K., {Desmond}, H., \& {Ferreira}, P.~G. 2019, \prd, 100, 123006

\bibitem[{{Percival} {et~al.}(2007){Percival}, {Nichol}, {Eisenstein},
  {Frieman}, {Fukugita}, {Loveday}, {Pope}, {Schneider}, {Szalay}, {Tegmark},
  {et~al.}}]{percival2007shape}
{Percival}, W.~J., {Nichol}, R.~C., {Eisenstein}, D.~J., {et~al.} 2007, \apj,
  657, 645

\bibitem[{{Planck Collaboration} {et~al.}(2020)}]{planck2018cosmo}
{Planck Collaboration} {et~al.} 2020, \aap, 641, A6

\bibitem[{{Porqueres} {et~al.}(2020){Porqueres}, {Hahn}, {Jasche}, \&
  {Lavaux}}]{porqueres2020hierarchical}
{Porqueres}, N., {Hahn}, O., {Jasche}, J., \& {Lavaux}, G. 2020, A\&A, 642,
  A139

\bibitem[{{Porqueres} {et~al.}(2021){Porqueres}, {Heavens}, {Mortlock}, \&
  {Lavaux}}]{porqueres2020shear}
{Porqueres}, N., {Heavens}, A., {Mortlock}, D., \& {Lavaux}, G. 2021, MNRAS,
  502, 3035

\bibitem[{{Porqueres} {et~al.}(2019{\natexlab{a}}){Porqueres}, {Jasche},
  {Lavaux}, \& {En{\ss}lin}}]{porqueres2019inferring}
{Porqueres}, N., {Jasche}, J., {Lavaux}, G., \& {En{\ss}lin}, T.
  2019{\natexlab{a}}, A\&A, 630, A151

\bibitem[{{Porqueres} {et~al.}(2019{\natexlab{b}}){Porqueres}, {Kodi Ramanah},
  {Jasche}, \& {Lavaux}}]{porqueres2019explicit}
{Porqueres}, N., {Kodi Ramanah}, D., {Jasche}, J., \& {Lavaux}, G.
  2019{\natexlab{b}}, A\&A, 624, A115

\bibitem[{{Porredon} {et~al.}(2021){Porredon}, {Crocce}, {Elvin-Poole},
  {Cawthon}, {Giannini}, {De Vicente}, {Carnero Rosell}, {Ferrero}, {Krause},
  {Fang}, {Prat}, {Rodriguez-Monroy}, {Pandey}, {Pocino}, {Castander}, {Choi},
  {Amon}, {Tutusaus}, {Dodelson}, {Sevilla-Noarbe}, {Fosalba}, {Gaztanaga},
  {Alarcon}, {Alves}, {Andrade-Oliveira}, {Baxter}, {Bechtol}, {Becker},
  {Bernstein}, {Blazek}, {Camacho}, {Campos}, {Carrasco Kind}, {Chintalapati},
  {Cordero}, {DeRose}, {Di Valentino}, {Doux}, {Eifler}, {Everett},
  {Fert{\'e}}, {Friedrich}, {Gatti}, {Gruen}, {Harrison}, {Hartley}, {Herner},
  {Huff}, {Huterer}, {Jain}, {Jarvis}, {Lee}, {Lemos}, {MacCrann},
  {Mena-Fern{\'a}ndez}, {Muir}, {Myles}, {Park}, {Raveri}, {Rosenfeld}, {Ross},
  {Rykoff}, {Samuroff}, {S{\'a}nchez}, {Sanchez}, {Sanchez}, {Sanchez Cid},
  {Scolnic}, {Secco}, {Sheldon}, {Troja}, {Troxel}, {Weaverdyck}, {Yanny},
  {Zuntz}, {Abbott}, {Aguena}, {Allam}, {Annis}, {Avila}, {Bacon}, {Bertin},
  {Bhargava}, {Brooks}, {Buckley-Geer}, {Burke}, {Carretero}, {Costanzi}, {da
  Costa}, {Pereira}, {Davis}, {Desai}, {Diehl}, {Dietrich}, {Doel},
  {Drlica-Wagner}, {Eckert}, {Evrard}, {Flaugher}, {Frieman},
  {Garc{\'\i}a-Bellido}, {Gerdes}, {Giannantonio}, {Gruendl}, {Gschwend},
  {Gutierrez}, {Hinton}, {Hollowood}, {Honscheid}, {Hoyle}, {James}, {Kuehn},
  {Kuropatkin}, {Lahav}, {Lidman}, {Lima}, {Lin}, {Maia}, {Marshall},
  {Martini}, {Melchior}, {Menanteau}, {Miquel}, {Mohr}, {Morgan}, {Ogando},
  {Palmese}, {Paz-Chinch{\'o}n}, {Petravick}, {Pieres}, {Plazas Malag{\'o}n},
  {Romer}, {Santiago}, {Scarpine}, {Schubnell}, {Serrano}, {Smith},
  {Soares-Santos}, {Suchyta}, {Tarle}, {Thomas}, {To}, {Varga}, \&
  {Weller}}]{porredon2021dark}
{Porredon}, A., {Crocce}, M., {Elvin-Poole}, J., {et~al.} 2021, ArXiv e-prints
  [\eprint[arXiv]{2105.13546}]

\bibitem[{{Racca} {et~al.}(2016){Racca}, {Laureijs}, {Stagnaro}, {Salvignol},
  {Lorenzo Alvarez}, {Saavedra Criado}, {Gaspar Venancio}, {Short}, {Strada},
  {B{\"o}nke}, {et~al.}}]{euclid2016missiondesign}
{Racca}, G.~D., {Laureijs}, R., {Stagnaro}, L., {et~al.} 2016, in Proc.~SPIE,
  Vol. 9904, Society of Photo-Optical Instrumentation Engineers (SPIE)
  Conference Series, 99040O

\bibitem[{{Saunders} {et~al.}(2000){Saunders}, {Sutherland}, {Maddox},
  {Keeble}, {Oliver}, {Rowan-Robinson}, {McMahon}, {Efstathiou}, {Tadros},
  {White}, {et~al.}}]{Saunders00}
{Saunders}, W., {Sutherland}, W.~J., {Maddox}, S.~J., {et~al.} 2000, MNRAS,
  317, 55

\bibitem[{{Schmidt} {et~al.}(2020){Schmidt}, {Cabass}, {Jasche}, \&
  {Lavaux}}]{schmidt2020unbiased}
{Schmidt}, F., {Cabass}, G., {Jasche}, J., \& {Lavaux}, G. 2020, J. Cosmology
  Astropart. Phys., 2020, 008

\bibitem[{{Skrutskie} {et~al.}(2006){Skrutskie}, {Cutri}, {Stiening},
  {Weinberg}, {Schneider}, {Carpenter}, {Beichman}, {Capps}, {Chester},
  {Elias}, {et~al.}}]{Skrutskie06}
{Skrutskie}, M.~F., {Cutri}, R.~M., {Stiening}, R., {et~al.} 2006, AJ, 131,
  1163

\bibitem[{{Tegmark} {et~al.}(2004)}]{tegmark2004threedimensional}
{Tegmark}, M. {et~al.} 2004, \apj, 606, 702

\bibitem[{{Villaescusa-Navarro} {et~al.}(2020){Villaescusa-Navarro},
  {Angl{\'e}s-Alc{\'a}zar}, {Genel}, {Spergel}, {Somerville}, {Dave},
  {Pillepich}, {Hernquist}, {Nelson}, {Torrey},
  {et~al.}}]{villaescusa2020camels}
{Villaescusa-Navarro}, F., {Angl{\'e}s-Alc{\'a}zar}, D., {Genel}, S., {et~al.}
  2020, ArXiv e-prints [\eprint[arXiv]{2010.00619}]

\bibitem[{{Villaescusa-Navarro} {et~al.}(2014){Villaescusa-Navarro}, {Marulli},
  {Viel}, {Branchini}, {Castorina}, {Sefusatti}, \& {Saito}}]{paco2014neutrino}
{Villaescusa-Navarro}, F., {Marulli}, F., {Viel}, M., {et~al.} 2014, J.
  Cosmology Astropart. Phys., 2014, 011

\bibitem[{{Wang} {et~al.}(2014){Wang}, {Mo}, {Yang}, {Jing}, \&
  {Lin}}]{wang2014ELUCID}
{Wang}, H., {Mo}, H.~J., {Yang}, X., {Jing}, Y.~P., \& {Lin}, W.~P. 2014, \apj,
  794, 94

\bibitem[{{Zhao} {et~al.}(2017){Zhao}, {Raveri}, {Pogosian}, {Wang},
  {Crittenden}, {Handley}, {Percival}, {Beutler}, {Brinkmann}, {Chuang},
  {et~al.}}]{zhao2017dynamical}
{Zhao}, G.-B., {Raveri}, M., {Pogosian}, L., {et~al.} 2017, Nature Astronomy,
  1, 627

\end{thebibliography}
%
% - join the .bib files when you upload your source files
%-------------------------------------------------------------------

\begin{appendix}
\label{appendix}
\section{2M++ galaxy catalog}
The 2M++ catalog \citep{LH11} is a compilation galaxy redshifts derived from the Two-Micron-All-Sky-Survey (2MASS) Redshift Survey (2MRS) \citep{Huchra12}, the Six-Degree-Field Galaxy Redshift Survey Data Release 3 (6dFGRS-DR3) \citep{Jones09}, and the Sloan Digital Sky Survey Data Release 7 (SDSS-DR7) \citep{SDSS7}. The resulting catalog has a greater depth and a higher sampling rate relative to the Infrared Astronomical Satellite (IRAS) Point Source Catalog Redshift Survey (PSCZ) \citep{Saunders00}. The photometry is based on the 2MASS Extended Source Catalog (2MASS-XSC) \citep{Skrutskie06}, an all-sky survey in the $J$, $H$ and $K_S$ bands, with redshifts in the $K_S$ band of the 2MRS complemented by those from the SDSS-DR7 and 6dFGRS-DR3.

Since the 2M++ catalog is a combination of several surveys, the galaxy magnitudes from all sources were first recomputed by measuring the apparent magnitude in the $K_S$ band within a circular isophote at 20 mags arcsec$^{-2}$. The apparent $K_S$ band magnitudes were subsequently corrected by taking into account Galactic extinction, cosmological surface brightness dimming, stellar evolution and $k$-corrections, while masking the Galactic Plane. To account for the incompleteness due to fibre collisions in 6dFGRS and SDSS, the redshifts of nearby galaxies within each survey region are cloned. The final 2M++ catalog contains 69190 galaxies in total, and is fairly sampling the galaxies of the cosmic volume up to a distance of 200\Mpch for the area covered by the 6dFGRS and SDSS, and up to 125\Mpch for the region mapped by 2MRS. For a more in-depth description of the construction of the 2M++ catalog, we refer the interested reader to the original compilation \citep{LH11}, with the computation of radial selection functions and target selection completeness as required for the \borg 2M++ analysis \citep{jasche2019physical} detailed in Section~4.1 thereof.

\section{BORG}
\label{app: appendixBORG}

The \borg (Bayesian Origin Reconstruction from Galaxies) algorithm \citep{jasche2013bayesian, jasche2019physical} constitutes a hierarchical Bayesian statistical inference framework to infer the initial conditions of the Universe from the observed galaxy positions in the sky as provided by galaxy redshift surveys. By encoding a physical forward model at its core, as illustrated in Fig.~\ref{fig:borg_schematic}, \borg allows the detailed 3D reconstruction and characterization of the observed cosmic large-scale structures.

The forward model entails the sequential application of several components that causally relate the cosmic initial conditions to the observed galaxy distribution, starting with a physical description of the non-linear dynamics involved in gravitational structure formation. While the \borg machinery incorporates several models of structure growth, such as the approximate first- and second-order Lagrangian perturbation theory and fully non-linear particle mesh models, we employ the latter option to evolve the primordial density fluctuations to their corresponding 3D dark matter distribution via cosmological $N$-body simulations. The predicted dark matter fields must subsequently be connected to a galaxy population via an adequate galaxy bias model. Although the galaxy biasing effect is a presently challenging and as yet unresolved issue in cosmology, the 2M++ analysis made use of a local but non-linear truncated power law bias model \citep{neyrinck2014halo} as motivated by numerical simulations. \borg inherently accounts for the physical effects such as the geometric distortion due to the cosmic expansion and the redshift space distortions due to the peculiar velocities of galaxies. The final component in the forward model deals with all the relevant observational and instrumental effects, such as survey geometry, selection effects and foreground contaminations, to yield the desired galaxy distribution that can be compared to observations via a suitable likelihood, with a Poissonian distribution adopted in the 2M++ analysis \citep{jasche2019physical}.

The above physical forward model results in a highly non-trivial Bayesian inverse problem. To efficiently sample the high-dimensional and non-linear parameter space of plausible initial conditions at an earlier epoch, with typically $\mathcal{O}(10^7)$ free parameters corresponding to the discretized volume elements of the observed domain, \borg relies on a Hamiltonian Monte Carlo (HMC) technique \citep{jasche2010fast, jasche2013bayesian, jasche2019physical}. \borg, in essence, performs the joint inference of the respective posterior distributions of the initial conditions and the bias parameters given the observed galaxy distribution via a modular Markov Chain Monte Carlo (MCMC) sampling framework. Feeding the MCMC samples of the initial conditions to the forward model therefore yields physically plausible realizations of the 3D cosmic structures, i.e. non-linearly evolved density fields and associated velocity fields, underlying the spatial distribution of observed galaxies.

Recent extensions to the \borg framework have led to substantial improvements in cosmological parameter inference \citep{DKR2018altair, elsner2020cosmology, schmidt2020unbiased}, Lyman-$\alpha$ \citep{porqueres2019inferring, porqueres2020hierarchical} and cosmic shear \citep{porqueres2020shear} reconstructions, with the field-level treatment transcending the capabilities of conventional cosmological analyses. Novel sophisticated additions to the forward model include a robust likelihood to account for unknown foreground contaminations and systematics \citep{porqueres2019explicit} and machine learning-based galaxy bias models \citep{charnock2020neural}. The reconstructed density and velocity fields from \borg analyses have been employed in investigating extensions of the standard model of particle physics and cosmology \citep{desmond2018fifth, desmond2019fifth, pardo2019testing, desmond2020galaxy, bartlett2021constraints}, and for unbiased and accurate measurements of the Hubble constant from gravitational wave events \citep{mukherjee2019velocity} as well as of the kinematic Sunyaev-Zel'dovich effect from CMB observations \citep{nguyen2020kSZ}.

\section{Fisher information maps}

The proposed targeted search approach requires identifying regions of the observational domain, which, when observed, can provide optimal information gain to update existing knowledge. To quantify this information gain, we choose to employ Fisher information \citep{fisher1925theory}. The latter measures the amount of information that unseen observable data can carry about uncertain model parameters. More explicitly, Fisher information estimates the squared norm of the likelihood score for given model parameters averaged over all possible future data realizations permitted by the likelihood. It, therefore, measures the expected strength with which the likelihood will respond to changes in the model parameters once new data becomes available. For the specific case considered in this work, we assume that prior knowledge on the spatial cosmic matter configuration is available from previous observations.
More specifically, we assume that the white noise realizations (the phases) of the initial conditions $\mvec{\phi}$ are provided by the \borg{} reconstruction of the 2M++ survey \citep{jasche2019physical}. Assuming this realization of initial conditions  we can express the conditional Fisher information as:
\begin{align*}
\mathcal{I}(\mvec{\theta}|\mvec{\phi}) &= \mathbb{E}_{(\mvec{d}|\mvec{\theta},\mvec{\phi})}\left[\left(\frac{\partial \ln(\mathcal{L}(\mvec{d}|\mvec{\theta},\mvec{\phi}))}{ \partial \mvec{\theta}} \right)^2\right] \\
&= \int \mathcal{D}{\mvec{d}} \, \left(\frac{\partial \ln(\mathcal{L}(\mvec{d}|\mvec{\theta},\mvec{\phi}))}{ \partial \mvec{\theta}} \right)^2  \mathcal{L}(\mvec{d}|\mvec{\theta},\mvec{\phi}) \, , \numberthis
\label{eqn: def_fisherinfo}
\end{align*}
where $\mvec{\theta}$ corresponds to the set of cosmological parameters parametrizing our forward model, with the galaxy observations within the voxels denoted by $\mvec{d} = [N_i]_{i=1,\dots N_{\rm{box}}}$ and $N_{\rm{box}}$ corresponds to the number of the voxels inside the 3D volume considered in the 2M++ \borg reconstruction \citep[see][for further details]{jasche2019physical}. Note that we express the vector nature of these random variables using boldface symbols. For the sake of this work, we assume a Poisson likelihood to describe the galaxy clustering data:
\begin{equation}
 \mathcal{L}(\mvec{d}|\mvec{\theta},\mvec{\phi}) = \prod_i \mathrm{e}^{-\lambda_i(\mvec{\theta};\mvec{\phi})} \frac{\lambda_i(\mvec{\theta};\mvec{\phi})^{N_i}}{N_i!} \, ,
\end{equation}
where $\lambda_i(\mvec{\theta};\mvec{\phi})$ denotes the rate of the specific realization of the Poisson process as dictated by the initial phases $\mvec{\phi}$ and cosmological parameters $\mvec{\theta}$, while $N_i$ represents the galaxy counts within the $i$-th cell of the 3D volume. Now, taking the derivative with respect to the cosmological parameters $\mvec{\theta}$ gives:
\begin{align*}
 \frac{\partial \ln(\mathcal{L}(\mvec{d}|\mvec{\theta},\mvec{\phi}))}{ \partial \mvec{\theta}} &= \sum_i \left( - \frac{\partial \lambda_i(\mvec{\theta};\mvec{\phi})}{ \partial \mvec{\theta}} + \frac{N_i}{\lambda_i(\mvec{\theta};\mvec{\phi})}\frac{\partial \lambda_i(\mvec{\theta};\mvec{\phi})}{ \partial \mvec{\theta}}   \right) \\
  &= \sum_i \left(  \frac{N_i}{\lambda_i(\mvec{\theta};\mvec{\phi})}-1   \right)  \frac{\partial \lambda_i(\mvec{\theta};\mvec{\phi})}{ \partial \mvec{\theta}} \, . \numberthis \\
\end{align*}
The square of the above derivative is given by:
\begin{align*}
\left(\frac{\partial \ln(\mathcal{L}(\mvec{d}|\mvec{\theta},\mvec{\phi}))}{ \partial \mvec{\theta}} \right)^2 &= \sum_{ij}  \frac{\partial \lambda_i(\mvec{\theta};\mvec{\phi})}{ \partial \mvec{\theta}} \frac{\partial \lambda_j(\mvec{\theta},\mvec{\phi})}{ \partial \mvec{\theta}} \nonumber \\
&\hspace{-25mm}\times \left( \frac{N_i}{\lambda_i(\mvec{\theta};\mvec{\phi})}\frac{N_j}{\lambda_j(\mvec{\theta},\mvec{\phi})}-\frac{N_i}{\lambda_i(\mvec{\theta};\mvec{\phi})}-\frac{N_j}{\lambda_j(\mvec{\theta},\mvec{\phi})} + 1  \right)  \nonumber \\
&\hspace{-25mm}= \sum_{i}  \left(\frac{\partial \lambda_i(\mvec{\theta};\mvec{\phi})}{ \partial \mvec{\theta}} \right)^2  \left( \frac{N_i^2}{\lambda_i^2(\mvec{\theta},\mvec{\phi})}-2\frac{N_i}{\lambda_i(\mvec{\theta};\mvec{\phi})} + 1     \right)  \\
&\hspace{-25mm}+ \sum_{ij} (1-\delta^K_{ij})  \frac{\partial \lambda_i(\mvec{\theta};\mvec{\phi})}{ \partial \mvec{\theta}} \frac{\partial \lambda_j(\mvec{\theta},\mvec{\phi})}{ \partial \mvec{\theta}} \\
&\hspace{-25mm}\times \left( \frac{N_i}{\lambda_i(\mvec{\theta};\mvec{\phi})}\frac{N_j}{\lambda_j(\mvec{\theta},\mvec{\phi})}-\frac{N_i}{\lambda_i(\mvec{\theta};\mvec{\phi})}-\frac{N_j}{\lambda_j(\mvec{\theta},\mvec{\phi})} + 1     \right) \, , \numberthis \label{eqn: fisher_split} 
\end{align*}
where we split the sum in diagonal and off-diagonal terms. Again, the indices $i$ and $j$ denote the pixels in the \textsc{healpix} projection. Since given a density field, the individual Poisson realizations are independent, the off-diagonal terms will vanish when computing the Fisher information as given by Eq.~\eqref{eqn: def_fisherinfo}. In order to see this, consider the following data average of Eq.~\eqref{eqn: fisher_split}:
\begin{multline*}
\left \langle \left( \frac {\partial \ln(\mathcal{L}(\mvec{d}|\mvec{\theta},\mvec{\phi}))} { \partial \mvec{\theta}} \right)^2 \right \rangle_{(\mvec{d} | \mvec{\theta}, \mvec{\phi})} = \sum_{ij}  \frac{\partial \lambda_i(\mvec{\theta};\mvec{\phi})}{ \partial \mvec{\theta}} \frac{\partial \lambda_j(\mvec{\theta};\mvec{\phi})}{ \partial \mvec{\theta}} \\
\times 
\left \langle \left( \frac{N_i}{\lambda_i(\mvec{\theta};\mvec{\phi})} \frac{N_j}{\lambda_j(\mvec{\theta},\mvec{\phi})} - \frac{N_i}{\lambda_i(\mvec{\theta};\mvec{\phi})} - \frac{N_j}{\lambda_j(\mvec{\theta};\mvec{\phi})} + 1 \right)
\right \rangle_{(\mvec{d}|\mvec{\theta}, \mvec{\phi})}\\
=\sum_{i} \left( \frac{\partial \lambda_i(\mvec{\theta};\mvec{\phi})}{\partial \mvec{\theta}} \right)^2 \left( \frac{\lambda_i(\mvec{\theta}; \mvec{\phi})^2 + \lambda_i(\mvec{\theta}; \mvec{\phi})}{\lambda_i^2(\mvec{\theta};\mvec{\phi})} - 2\frac{\lambda_i(\mvec{\theta}; \mvec{\phi})}{\lambda_i(\mvec{\theta};\mvec{\phi})} + 1 \right)  \nonumber \\
+ \sum_{ij} (1-\delta^K_{ij})  \frac{\partial \lambda_i(\mvec{\theta};\mvec{\phi})}{ \partial \mvec{\theta}} \frac{\partial \lambda_j(\mvec{\theta};\mvec{\phi})}{ \partial \mvec{\theta}} \nonumber \\
\underbrace{ \left( \frac{\lambda_i(\mvec{\theta}; \mvec{\phi})}{\lambda_i(\mvec{\theta};\mvec{\phi})} \frac{\lambda_j(\mvec{\theta}; \mvec{\phi})}{\lambda_j(\mvec{\theta},\mvec{\phi})} - \frac{\lambda_i(\mvec{\theta}; \mvec{\phi})}{\lambda_i(\mvec{\theta};\mvec{\phi})} - \frac{\lambda_j(\mvec{\theta}; \mvec{\phi})}{\lambda_j(\mvec{\theta};\mvec{\phi})} + 1 \right)}_{=0} \, ,
\numberthis \label{eqn: fisher_d_averaged}
\end{multline*}
where we used the following identities:
\begin{align*}
\langle N_i \rangle_{(\mvec{d} | \mvec{\theta}, \mvec{\phi})} &= \lambda_i(\mvec{\theta}; \mvec{\phi}),\\
\langle N_i^2 \rangle_{(\mvec{d} | \mvec{\theta}, \mvec{\phi})} &= \lambda_i(\mvec{\theta}; \mvec{\phi}) + \lambda_i^2(\mvec{\theta}; \mvec{\phi}) \hspace{3mm}  \text{for $\forall i \in [1, N]$}.
\numberthis \label{eqn: identities}
\end{align*}
The first equality follows from the conditional independence of the Poisson realizations within each grid cell, which always holds once the underlying density field $\mvec{\lambda}(\mvec{\theta}, \mvec{\phi})$ is given, while the second results from the expression of variance of the Poisson distribution.
Therefore, the diagonal terms are:
\begin{equation}
\mathcal{I}(\mvec{\theta} | \mvec{\phi}) = \sum_{i}  \left(\frac{\partial \lambda_i(\mvec{\theta};\mvec{\phi})}{ \partial \mvec{\theta}} \right)^2 \frac{1}{\lambda_i(\mvec{\theta};\mvec{\phi})} \, .  
\end{equation}
From the above equation, we see that the contribution from a particular volume element to the Fisher information is given by:
\begin{equation}
h'_i (\mvec{\theta} | \mvec{\phi}) = 
\left(\frac{\partial \lambda_i(\mvec{\theta};\mvec{\phi})}{ \partial \mvec{\theta}} \right)^2 \frac{1}{\lambda_i(\mvec{\theta};\mvec{\phi})} \, 
\equiv
\lambda_i(\mvec{\theta}, \mvec{\phi}) 
\left(\frac{\partial \ln \lambda_i(\mvec{\theta}; \mvec{\phi})}{\partial \mvec{\theta}} \right)^2 \, . 
\numberthis
\label{eqn: h_i_expression_1}
\end{equation}
This allows us to obtain a 3D Fisher map. Now, we also have to specify the Poisson intensity in terms of the output of our physics simulator. This relation can be written as:
\begin{equation}
\lambda_i(\mvec{\theta};\mvec{\phi}) = B(\delta \equiv G(\mvec{\theta};\mvec{\phi})) \, ,
\end{equation}
where $B(x)$ is an arbitrary non-negative bias function and $G(\mvec{\theta},\mvec{\phi})$ is a physics simulator of the cosmic large-scale structures with output $\delta$ corresponding to the 3D matter density contrast amplitudes. For illustrative purposes, we assume $B(x)= 1 + x$. Note any bias model monotonic in the density will change the quantitative results, but not the qualitative results, as can also be seen from Eq.~\eqref{eqn: h_i_expression_1}. We then obtain:
\begin{equation}
\lambda_i(\mvec{\theta};\mvec{\phi}) = 1 + G_i(\mvec{\theta};\mvec{\phi}) \, ,
\end{equation}
which leads to the following expression for the derivative with respect to cosmological parameters $\mvec{\theta}$:
\begin{equation}
\frac{\partial \lambda_i(\mvec{\theta};\mvec{\phi})}{ \partial \mvec{\theta}} =  \frac{\partial G_i(\mvec{\theta};\mvec{\phi})}{ \partial \mvec{\theta}} \, .
\end{equation}
We can therefore approximate the Fisher information elements through finite differencing as:
\begin{align*}
h'_i(\mvec{\theta};\mvec{\phi}) &= \left(\frac{\partial G_i(\mvec{\theta};\mvec{\phi})}{ \partial \mvec{\theta}} \right)^2   \frac{1}{1 + G_i(\mvec{\theta};\mvec{\phi})} \\
&\approx \left(\frac{ G_i(\mvec{\theta}_0 + \Delta \mvec{\theta};\mvec{\phi})-G_i(\mvec{\theta}_0;\mvec{\phi})}{\Delta \mvec{\theta}} \right)^2 \frac{1}{1 + G_i(\mvec{\theta}_0;\mvec{\phi})} \, . 
\numberthis  \label{eqn: finite_difference_maps_fisher}  
\end{align*}
Note, however, that the above calculation is done for a fixed realization of the phases $\mvec{\phi}$. Since we do not know what particular realization is compatible with our Universe, we need to marginalize over them:

\begin{align*}
h_i (\mvec{\theta} | d) &=
\int \mathcal{D} \mvec{\phi} \,
p(\mvec{\phi} | d)
h'_i(\mvec{\theta}|\mvec{\phi}) \\
&\equiv
\int \mathcal{D} \mvec{\phi} \,
p(\mvec{\phi} | d)
\left(\frac{\partial G_i(\mvec{\theta};\mvec{\phi})}{ \partial \mvec{\theta}} \right)^2   \frac{1}{1 + G_i(\mvec{\theta};\mvec{\phi})} .
\numberthis 
\label{eqn: h_i_expression}
\end{align*}

This task is made possible by the \borg algorithm, as described in Appendix~\ref{app: appendixBORG}. In this way, we are able to use the information content of the previously obtained constraints and update the Fisher information. Projecting this 3D Fisher map onto a \textsc{healpix} grid and visualizing a particular spherical slice through it, we obtain an all-sky map which we refer to in our study as the {\it Fisher information map}. A particular example is provided in Fig.~\ref{fig:fisher_maps_PM}.

Now, in order to calculate the $G_i(\mvec{\theta}_0 + \Delta \mvec{\theta}; \mvec{\phi})$ term above, we first perform a forward model evaluation by setting the cosmological parameters to their corresponding fiducial values, $\mvec{\theta}_0 \equiv \{ \Omega_{\mathrm{m}}=0.3111, \Omega_{\mathrm{b}}=0.049, \Omega_{\mathrm{k}} = 7 \times 10^{-4}, h=0.6766, \sigma_8=0.8102, n_{\mathrm{s}}=0.9665 \}$, according to the latest Planck best-fit $\Lambda$CDM cosmology \citep{planck2018cosmo}, for all the MCMC realizations of the initial conditions from the \borg 2M++ analysis. We then compute the mean galaxy field marginalized over the forward model output realizations. We subsequently repeat this procedure for a new set of perturbed cosmological parameters, $\mvec{\theta}' \equiv \mvec{\theta}_0 + \Delta \mvec{\theta}$, ensuring $\Delta \mvec{\theta}$ does not exceed values outside 1-sigma width of a Gaussian centered at the fiducial cosmology $\mvec{\theta}_0$ as characterized by Planck best-fit values. Using the respective means of the fiducial and perturbed galaxy fields, we compute the gradient of the forward model output with respect to the cosmological parameters, $\Delta_{G}/\Delta_{\mvec{\theta}}$, using a finite differencing scheme, with:

\begin{align*}
\Delta_{G} &= \int \mathcal{D}\mvec{\phi} \, p(\mvec{\phi} | d) \left[G(\mvec{\theta}_0 + \Delta \mvec{\theta}; \mvec{\phi}) - G(\mvec{\theta}_0;\mvec{\phi})\right]\\
&\approx \frac{1}{N_{\mvec{\phi}}}
\sum_i
\left[G(\mvec{\theta}_0 + \Delta \mvec{\theta}; \mvec{\phi}_i) - G(\mvec{\theta}_0;\mvec{\phi}_i)\right]
\hspace{5mm} \text{with } \mvec{\phi}_i \hookleftarrow p(\mvec{\phi}|\mvec{d}) 
\end{align*} 
and $\Delta \mvec{\theta} = \mvec{\theta}' - \mvec{\theta}_0$. The distinct gradient components correspond to the cosmological sensitivity maps displayed in Fig.~\ref{fig:dmaps_PM}. The gradient is then squared and divided by the mean fiducial galaxy field and marginalized over phase realizations, as required by Eq.~\eqref{eqn: finite_difference_maps_fisher}, to finally yield the desired Fisher information elements. The resulting Fisher information map for a particular spherical slice through the 3D field is illustrated in Fig.~\ref{fig:fisher_maps_PM}.

In this approach, we do not explicitly enforce that $\sum_i \Omega_i = 1$ for $\mvec{\theta}'$, since we are interested in infinitesimal and independent variations of the cosmological parameters as required by the Fisher information. Therefore, although this is marginally inconsistent from the cosmological perspective, it is perfectly consistent from the information theory perspective. Furthermore, enforcing the $\sum_i \Omega_i = 1$ in our forward model introduces dependencies of the variations required for the evaluation of the Fisher information, which renders the interpretation more convoluted. Nonetheless, we also tested the impact of enforcing $\sum_i \Omega_i = 1$ in our forward model and found that this has a negligible effect on the derived Fisher information map. In addition, we verified the robustness of our results, as expected, to the choice of fiducial cosmology.

\section{Marginalizing phase realizations and targeted searches}
\label{app: appendixD}

The derivation presented in the previous section took care of explicitly keeping, where needed, the conditional dependence on the specific realization of the phases of the initial density field, also referred to as the initial white noise field. The reason being that in practice, the exact phase information of the initial density field is not available:
\begin{equation}
\mathcal{I}(\mvec{\theta}) = \int \mathrm{d}\mvec{\phi} \, \mathcal{P}(\mvec{\phi}) \,  \mathcal{I}(\mvec{\theta}|\mvec{\phi}) \, ,
\end{equation}
where $\mathcal{P}(\mvec{\phi})$ is a Gaussian prior with zero mean and unit variance for the initial white noise field. We effectively use a Markov approximation of the integral by drawing random realizations from the white noise prior and evaluating the corresponding averaged Fisher information as:
\begin{equation}
\mathcal{I}(\mvec{\theta}) = \frac{1}{N_{\mvec{\phi}}} \sum_i  \mathcal{I}(\mvec{\theta}|\mvec{\phi}_i) \, .
\end{equation}
where the $\mvec{\phi}_i$ are independent white noise realizations. Similarly, we can just decompose the averaged Fisher information into the individual components per volume element:
\begin{equation}
h_j(\mvec{\theta}) = \frac{1}{N_{\mvec{\phi}}} \sum_i  h_j(\mvec{\theta}|\mvec{\phi}_i) \, .
\end{equation}
It is clear that since we marginalize over all possible white noise realizations, there cannot be any variation in the $h_i(\mvec{\theta})$ since every configuration is equally likely. Thus, without any further prior information on the phases, every point in the Universe, on average, is expected to contribute exactly the same amount of information.

This is one of the logical reasons behind performing homogeneous cosmological surveys. However, we are no longer in a regime of cosmology where complete ignorance about the Universe prevails. We have access to data, which can inform us about the specific realization of our Universe. If we account for this fact, we can condition the average Fisher information on the existing data:
\begin{equation}
\mathcal{I}(\mvec{\theta}|\mvec{d}) = \int \mathrm{d}\mvec{\phi} \, \mathcal{P}(\mvec{\phi}|\mvec{d}) \,  \mathcal{I}(\mvec{\theta}|\mvec{\phi}) \, .     
\end{equation}
Now, $\mathcal{P}(\mvec{\phi}|\mvec{d})$ is the data constrained posterior of the initial phase distribution. Obtaining this posterior distribution is a non-trivial task, but one that is solved by our \borg algorithm. In particular, \borg provides a Markov approximation to the high-dimensional posterior distribution $\mathcal{P}(\mvec{\phi}|\mvec{d})$, such that we can approximate the integral as:
\begin{equation}
\mathcal{I}(\mvec{\theta}|\mvec{d}) = \frac{1}{N_{\mvec{\phi}}^{\borg}} \sum_i  \mathcal{I}(\mvec{\theta}|\mvec{\phi}^{\borg}_i) \, ,  
\end{equation}
with this inferred posterior of initial phase distributions being fairly robust to the details of the physical model adopted \citep{nguyen2021impacts}. Similarly, we can use the results of \borg to estimate the Fisher information per volume element as:
\begin{equation}
h_j(\mvec{\theta}|\mvec{d}) = \frac{1}{N_{\mvec{\phi}}^{\borg}} \sum_i  h_j(\mvec{\theta}|\mvec{\phi}^{\borg}_i) \, .   
\numberthis \label{eqn: phasemarg_fishervoxel}
\end{equation} 
The key point to realize here is that the result does not depend on the white noise phases $\mvec{\phi}$ as they have been marginalized out. More precisely, this has been achieved by evaluating Eq.~\eqref{eqn: phasemarg_fishervoxel} with $N_{\mvec{\phi}}^{\borg{}} = 2422$ samples characterizing the posterior $\mathcal{P}(\mvec{\phi} | \mvec{d})$. \par

\end{appendix}

\end{document}